\documentclass[letterpaper,twocolumn,10pt]{article}
\usepackage{usenix-2020-09}
\usepackage[available]{usenixbadges}

\usepackage{graphicx} % Required for inserting images
\usepackage[utf8]{inputenc}
\usepackage{tikz}
\usepackage{xspace}
\usepackage{xcolor}
\usepackage{xstring}
\usepackage{color, colortbl}
\usepackage{multicol,multirow}
\let\labelindent\relax
\usepackage{enumitem}
\usepackage{booktabs}
\usepackage{subcaption}
\usepackage{fontawesome}
\usepackage{amsmath}
\usepackage{bm}
\usepackage{nicematrix}
\usepackage[pangram]{blindtext}
\usepackage{balance}
\usepackage{totpages}
\usepackage[numbers,sort&compress]{natbib}
\usepackage{listings}
\usepackage[normalem]{ulem}
\usepackage{multirow}
\usepackage{booktabs}
\usepackage{makecell}
\usepackage{enumitem}
\usepackage{array} 
\usepackage{seqsplit}
\usepackage{float}
\usepackage{diagbox}
\usepackage{threeparttable}
\usepackage{hyperref}
\usepackage{cleveref}
\usepackage{microtype}

\usepackage[framemethod=tikz]{mdframed}
\usepackage{tabularx}
\usepackage{tcolorbox}

\usepackage{authblk}

\newcommand{\etal}{\textit{et al}.\xspace}
\newcommand{\ie}{\textit{i}.\textit{e}.}
\newcommand{\eg}{\textit{e}.\textit{g}.}

\def\Snospace~{\S{}}

\if 0

\fi

\newif\ifdraft\drafttrue
\newif\ifnotes\notestrue
\ifdraft\else\notesfalse\fi

\input{glyphtounicode}
\newcommand{\squishlist}{
\begin{itemize}[noitemsep,nolistsep]
  \setlength{\itemsep}{-0pt}
}
\newcommand{\squishend}{
  \end{itemize}
}

\setlist[itemize]{leftmargin=.15in, topsep={2pt}, partopsep={0pt}}

\DeclareRobustCommand*\BC[1]{%
\begin{tikzpicture}[baseline=(C.base)]
\node[draw,circle,fill=black,inner sep=0.2pt](C) {\textcolor{white}{#1}};
\end{tikzpicture}}

\newcommand{\PP}[1]{
\vspace{2px}
\noindent{\bf \IfEndWith{#1}{.}{#1}{#1.}}
}

\newcommand{\PPNS}[1]{
\noindent{\bf \IfEndWith{#1}{.}{#1}{#1.}}
}

\newcommand{\boxbeg}{
\vspace{2px}
\noindent\begin{tabular}{|l|}\hline
\begin{minipage}{3.2in}
\vspace{2px}
\noindent
}

\newcommand{\boxend}{
\vspace{2px}
\end{minipage}\\ \hline
\end{tabular}
\vspace{-10pt}
}

\newcommand{\reducedstrut}{\vrule width 0pt height 1.1\ht\strutbox depth 0.8\dp\strutbox\relax}

\newcommand{\aref}[1]{\hyperref[#1]{Appendix~\ref*{#1}}}

\newcommand*\justify{%
  \fontdimen2\font=0.4em% interword space
  \fontdimen3\font=0.2em% interword stretch
  \fontdimen4\font=0.1em% interword shrink
  \fontdimen7\font=0.1em% extra space
  \hyphenchar\font=`\-% allowing hyphenation
}

\renewcommand{\texttt}[1]{%
  \begingroup
  \ttfamily
  \begingroup\lccode`~=`/\lowercase{\endgroup\def~}{/\discretionary{}{}{}}%
  \begingroup\lccode`~=`[\lowercase{\endgroup\def~}{[\discretionary{}{}{}}%
  \begingroup\lccode`~=`.\lowercase{\endgroup\def~}{.\discretionary{}{}{}}%
  \catcode`/=\active\catcode`[=\active\catcode`.=\active
  \justify\scantokens{#1\noexpand}%
  \endgroup
}

\newcolumntype{P}[1]{>{\centering\arraybackslash}p{#1}}

\newtcolorbox{boxI}{
    colback = lightgray!10, 
    colframe = black, 
    boxrule = 0.5pt, 
    toprule = 0.5pt, % top rule weight
    arc = 2pt,
    left = 1pt,
    right = 1pt,
    bottom = 0pt,
    top = 0pt
}

\newcounter{observcntr}
\newcommand*{\observ}[1]{%
    \stepcounter{observcntr}%
    \begin{center}
    \vspace{-10px}
        \begin{boxI}
        \textbf{Takeaway~\arabic{observcntr}: }{#1}.
        \end{boxI}
    \vspace{-8px}    
    \end{center}
}

\RequirePackage{color}\definecolor{RED}{rgb}{1,0,0}\definecolor{BLUE}{rgb}{0,0,1} %DIF PREAMBLE
\begin{document}
%-------------------------------------------------------------------------------

%don't want date printed
\date{}

\title{\Large \bf Evaluating the Effectiveness and Robustness of \\ Visual Similarity-based Phishing Detection Models}

\author{Fujiao Ji$^1$, Kiho Lee$^1$, Hyungjoon Koo$^2$, Wenhao You$^3$, Euijin Choo$^3$, Hyoungshick Kim$^2$, Doowon Kim$^{1}$}
\affil{\textit{$^1$University of Tennessee, Knoxville\hspace{0.5em} $^2$Sungkyunkwan University\hspace{0.5em} $^3$University of Alberta}}

\maketitle

%%%%%%%%full version%%%%%%%%%%%
\begin{abstract}
Phishing attacks pose a significant threat to Internet users, with cybercriminals elaborately replicating the visual appearance of legitimate websites to deceive victims. Visual similarity-based detection systems have emerged as an effective countermeasure, but their effectiveness and robustness in real-world scenarios have been underexplored. In this paper, we comprehensively scrutinize and evaluate the effectiveness and robustness of popular visual similarity-based anti-phishing models using a large-scale dataset of 451k real-world phishing websites. Our analyses of the effectiveness reveal that while certain visual similarity-based models achieve high accuracy on curated datasets in the experimental settings, they exhibit notably low performance on real-world datasets, highlighting the importance of real-world evaluation. Furthermore, we find that the attackers evade the detectors mainly in three ways: (1) directly attacking the model pipelines, (2) mimicking benign logos, and (3) employing relatively simple strategies such as eliminating logos from screenshots. To statistically assess the resilience and robustness of existing models against adversarial attacks, we categorize the strategies attackers employ into visible and perturbation-based manipulations and apply them to website logos. We then evaluate the models' robustness using these adversarial samples. Our findings reveal potential vulnerabilities in several models, emphasizing the need for more robust visual similarity techniques capable of withstanding sophisticated evasion attempts. We provide actionable insights for enhancing the security of phishing defense systems, encouraging proactive actions. 
\end{abstract}

\vspace{-5px}
\section{Introduction} 
\label{sec:introduction}
Phishing attacks threaten Internet users' security through deceptive websites that mimic legitimate ones~\cite{verizon2021,ho2019detecting}.
Attackers replicate authentic sites of financial services or social media (\eg, PayPal, Facebook), copying visual elements (\eg, logos and layouts) to trick users into revealing sensitive credentials.
In the ongoing battle against phishing attacks, anti-phishing systems employ multiple detection strategies. These defensive measures examine URLs~\cite{Kim2022URL, Thirumuruganathan2022SP, Lee2021EuroSP}, HTML structure~\cite{Gur2023Understanding, Montaruli2023Raze, Opara2020HTMLPhish}, and visual elements~\cite{Lin2021phishpedia,Liu2022phishintention,Abdelnabi2020visualphishnet,Fu2006EMD,Liu2006anti,Afroz2011PhishZoo, Dynaphish2023Liu} to identify fraudulent websites. The visual components (\eg, logos and layouts) of websites have proven particularly critical in the phishing landscape, as attackers primarily rely on visual deception to establish credibility with potential victims.
In response, visual similarity-based detection models have become an essential component of modern anti-phishing defenses, using deep learning techniques to identify fraudulent sites that closely resemble well-known target brands.

\begin{figure}[!t]
    \centering
    \begin{subfigure}[t]{0.49\linewidth}
        \includegraphics[width=\linewidth]{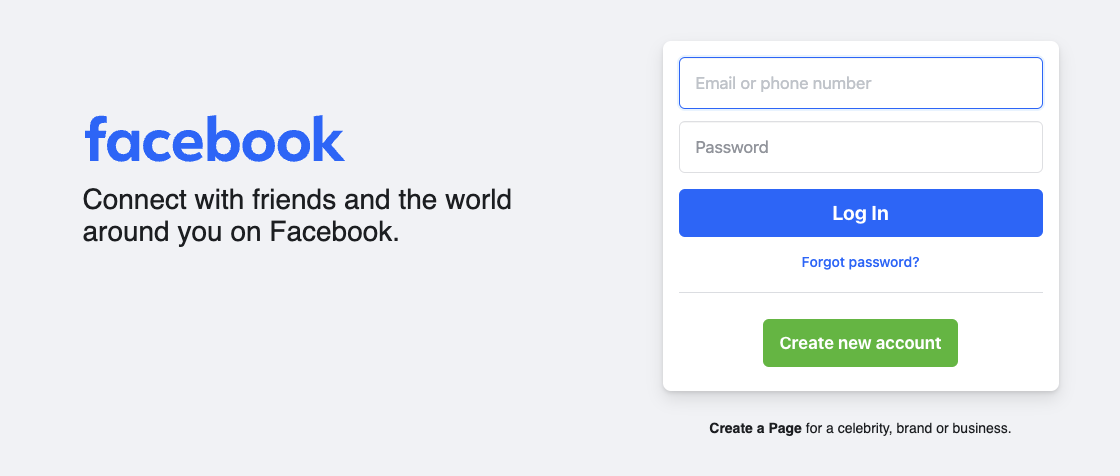} % 
        \caption{Original Login Form of \url{facebook.com}}
        \label{fig:first_image}
    \end{subfigure}
    \hfill 
    \begin{subfigure}[t]{0.49\linewidth}
        \includegraphics[width=\linewidth]{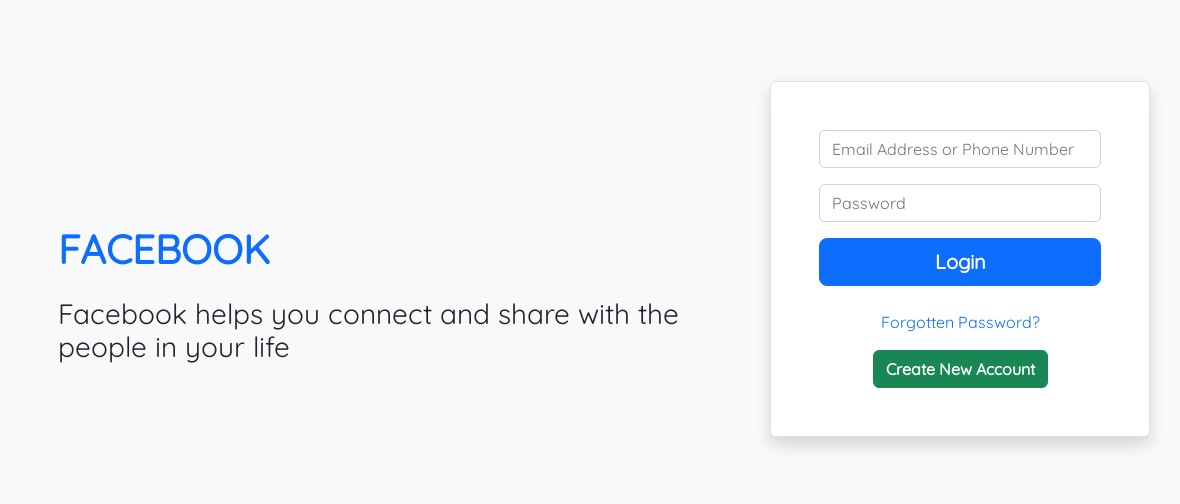} 
        \caption{Adversarial Manipulation of Logo Text  (Upper Case and Font)}
        \label{fig:second_image}
    \end{subfigure}
    \vspace{-5px}
    \caption{\textbf{Examples of Original Login Form and Adversarial Manipulation.} An attacker changes the textual logo (`facebook') to its upper case (`FACEBOOK') and its font.  The (b) example is found in our real-world phishing dataset.}~\label{fig:facebook_attack_example}
    \vspace{-20px}
\end{figure}

% In the ongoing battle against phishing attacks, anti-phishing ecosystems continually strive to thwart these malicious activities by leveraging diverse phishing patterns and advanced analytical techniques, such as URL characteristics~\cite{Kim2022URL, Thirumuruganathan2022SP, Lee2021EuroSP} and HTML source code~\cite{Gur2023Understanding, Montaruli2023Raze, Opara2020HTMLPhish}. However, URLs alone often provide insufficient information for training classifiers, and attackers can employ cloaking techniques~\cite{PhishDecloaker2024Teoh} to bypass HTML-based models~\cite{Divakaran2022Review}. 
% Thus, the visual elements of websites have played a crucial role in detecting phishing websites. 
% Visual similarity-based defense mechanisms~\cite{Lin2021phishpedia,Liu2022phishintention,Abdelnabi2020visualphishnet,Fu2006EMD,Liu2006anti,Afroz2011PhishZoo, Dynaphish2023Liu} usually employ deep learning neural networks to compare and analyze visual characteristics using screenshots or target brand logos.
% These mechanisms flag potential phishing attempts by identifying fraudulent sites that closely resemble well-known target brands (\eg, Facebook) but operate under different domains.

Prior works~\cite{Look2024Hao,Thomas2020Phishing, Liu2022phishintention} analyzed the robustness of phishing detectors. 
Particularly, Hao \textit{et al.}~\cite{Look2024Hao} evaluated the robustness of detection models against their perturbation attacks where logo images are perturbed while preserving their semantic meaning.
They found potential weaknesses in existing detection models.
However, their work explored only limited perturbation techniques.
Moreover, prior works evaluated their models without considering different conditions (\eg datasets).
Therefore, there are three major limitations: the lack of (1) systematic evaluations assessing the effectiveness and robustness of multiple detectors under consistent, fair, and large-scale real-world conditions, (2) in-depth analyses of influential factors in adversarial attacks (\eg,~\autoref{fig:facebook_attack_example}), and their impact on detection failures, and (3) efforts to identify specific weaknesses associated with each influential factor. These limitations impede the development of more actionable and concrete recommendations for enhancing these approaches.

Our work addresses these gaps in visual similarity-based phishing detection through comprehensive evaluations of prominent models using large-scale real-world datasets. By analyzing factors influencing adversarial attack outcomes, including image manipulation, layout changes, and color alterations, we systematically identify and categorize model weaknesses for each influential factor. This approach yields actionable recommendations for improving detection methods, guided by two key research questions: \textbf{RQ1:} Do visual similarity-based anti-phishing mechanisms maintain their \textit{effectiveness} and \textit{robustness} against real-world phishing attacks under the same experimental settings? \textbf{RQ2:} Are visual similarity-based anti-phishing mechanisms sufficiently resilient against adversarial strategies that manipulate visual components to evade detection?

In response to RQ1, we conduct a comprehensive performance evaluation of popular models on a large-scale dataset comprising 451k real-world phishing websites, 4,190 sampled phishing websites, and 2,500 benign samples (of Tranco Top 1000 websites (\url{https://tranco-list.eu/})), to dig out the potential factors that are influential to the performance in phishing detection. 
While \texttt{PhishZoo}~\cite{Afroz2011PhishZoo} initially appears promising with high detection accuracy (78.25\%), our deeper analysis reveals significant limitations, including an elevated false positive rate (93.2\%) and poor brand identification capabilities (12.78\%).
This indicates that these severe deficiencies may render the model impractical for real-world deployment.

We find that other models exhibit significantly lower performance compared to their original reported results on curated datasets. This discrepancy can be attributed to multiple factors, such as model structures and dataset attributes, highlighting the importance of evaluating models on real-world data to assess their actual performance. 
Furthermore, our study reveals that static brand reference lists used for brand-domain matching in \texttt{PhishIntention}~\cite{Liu2022phishintention} and \texttt{Phishpedia}~\cite{Lin2021phishpedia} can be limited in real-world scenarios where websites regularly update their layouts and rebrand their logos. 
We also identify that attackers may craft phishing websites to directly attack the model pipeline, mimic legitimate websites, and use relatively simple strategies based on our analysis of failed examples. For example, simply eliminating logos will lead to the failures of logo-based methods because they can not recognize the brands of phishing webpages to verify the brand and domain.
\looseness=-1

To address RQ2, we manually analyze 6,000 detection failures to quantify key strategies attackers might employ. We then test these strategies using data from 110 popular benign websites, applying various visible and adversarial perturbations to visual components, particularly logos, to evaluate model resilience. Our findings reveal that both simple and adversarial manipulations can significantly undermine logo-based detection methods. These adversarial attacks are transferable across detection models. Although screenshot-based methods maintain stable detection, they struggle with accurately identifying brands when logos are altered. This evaluation offers crucial insights for developing more resilient models against adversarial attacks and evasion tactics.
\looseness=-1

The following summarizes our contributions.

\begin{itemize}[leftmargin=*, topsep=0pt, itemsep=0pt]
  \item We conduct the first comprehensive study using a large-scale dataset of over 451k real-world phishing websites to fairly evaluate seven visual similarity-based anti-phishing systems by ensuring systems know the same brand knowledge. Our findings suggest that these systems are less effective in real-world scenarios, indicating significant performance degradation (20.7\%), compared to their results on curated datasets. 
  
  \item We also find three ways attackers usually employ to bypass detectors: (1) exploiting weaknesses of models' pipelines (\eg, removing login forms), (2) mimicking benign logos and screenshots in the feature space, and (3) relatively simple strategies (\eg, changing colors of logos).
  
  \item For robustness evaluation, we show critical limitations in visual similarity-based phishing detection models against adversarial samples.

  \item Based on our findings, we recommend several strategies to improve the effectiveness and robustness of visual similarity-based anti-phishing mechanisms. These include integrating text recognition with visual analysis and using preprocessing techniques such as scaling and denoising to minimize the impacts of adversarial perturbations.

  \item We publicly share our collected real-world phishing dataset, our manipulated dataset, code, and re-trained models at our website~\url{https://moa-lab.net/evaluation-visual-similarity-based-phishing-detection-models/}.

\end{itemize}
\vspace{-5px}
\section{Background} \label{sec:background}

\PP{Phishing}
Phishing is a type of social engineering attack in which attackers try to trick victims into disclosing sensitive information (\eg, credentials). 
A phishing campaign involves fraudulent websites that mimic the appearance of legitimate websites. 
Victims are lured into disclosing their sensitive information to the attackers. Typically, such stolen information could be misused for further fraud or crimes.

\PP{Visual Similarity-based Phishing Detection Systems}
URL-based phishing detection systems primarily rely on blocklist-based defense mechanisms (\eg, Google Safe Browsing~\cite{SafeBrow50:online}) or machine learning models (\eg, \cite{Le2018URLNet, Woodbridge2018IEEESandP, Li2019Domain}) to prevent users from accessing malicious websites. However, relying solely on URLs is insufficient, as they provide limited information about a website's content, structure, or visual appearance, which are crucial for accurate phishing detection.
To address these limitations, research has focused on analyzing visual components of phishing websites, such as screenshots and target brand logos. 
Early approaches used Earth Mover's Distance~\cite{Fu2006EMD,hitchcock1941distribution}, and SIFT~\cite{Afroz2011PhishZoo} for image matching, and assessments of block, layout, and style similarities~\cite{Liu2006anti}. Recent deep learning advancements have introduced more sophisticated methods. \texttt{VisualPhishNet}~\cite{Abdelnabi2020visualphishnet} uses triplet CNNs for learning visual similarities between webpage screenshots, while \texttt{Phishpedia} and \texttt{PhishIntention} combine Faster-RCNN~\cite{Ren2015fastrcnn} for logo recognition with a Siamese architecture for similarity comparison. \texttt{DynaPhish}~\cite{Dynaphish2023Liu} utilizes Google search to identify targeted brands and dynamically expand the reference lists.
\looseness=-1

\begin{figure*}[!t]
\vspace{-5px}
\centering
\includegraphics[width=1\textwidth]{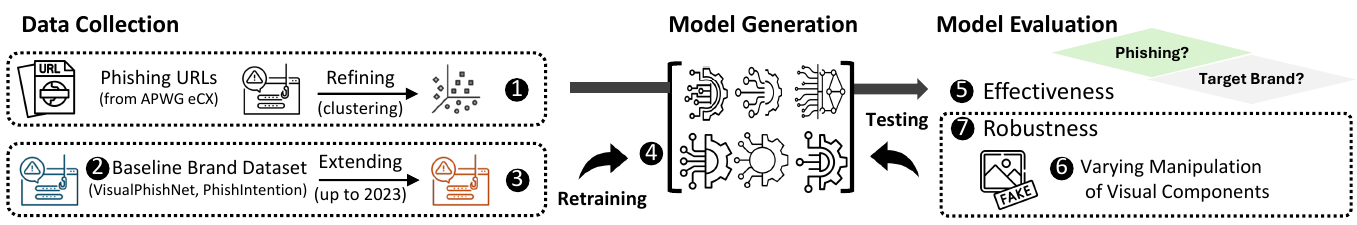}
\vspace{-10px}
\caption{\textbf{Overview of Our Experiment.} 
We collect real-world phishing and benign websites (\protect{\BC{1}}).
We prepare two reference datasets (\protect{\BC{2} and \BC{3}}).
We then carefully select seven popular visual similarity-based anti-phishing models and re-train them using the prepared datasets (\protect{\BC{4}}).
The effectiveness and robustness of these models are systematically evaluated (\protect{\BC{5}}, \protect{\BC{6}}, and \protect{\BC{7}}).
%%%
% Our experiment begins with collecting a real-world phishing and benign dataset (\ie, screenshots and HTML files) using URLs from APWG \texttt{eCX} and Tranco 1M domains. The phishing dataset is refined by filtering out pages involving errors (\eg, HTTP 404) or CAPTCHA through clustering (\protect{\BC{1}}; \autoref{sec:dataset:collection}). The APWG dataset is used solely for testing purposes. In addition, we prepare two brand reference datasets: a baseline dataset combining PhishIntention~\cite{Liu2022phishintention} and VisualPhishNet~\cite{Abdelnabi2020visualphishnet}, and extended datasets with more reference logos or screenshots (\protect{\BC{2} and \BC{3}}; See~\autoref{tab:dataset}). We then carefully select seven popular visual similarity-based anti-phishing models~\cite{Fu2006EMD, Afroz2011PhishZoo, Abdelnabi2020visualphishnet, Li2021CVPR, Lin2021phishpedia, Liu2022phishintention, Dynaphish2023Liu} and re-train them using the prepared datasets (\protect{\BC{4}}; \autoref{sec:selection:models}). The effectiveness of these models is systematically evaluated (\protect{\BC{5}}; \autoref{sec:retraining:models}) using our refined APWG dataset and benign dataset. To further assess the models, we manipulate visual components (\eg, logo images) with varying transformations (\protect{\BC{6}}; \autoref{sec:design:manipulating}) and quantify the robustness of the models against these manipulations (\protect{\BC{7}}).
}~\label{fig:overview}
\vspace{-10px}
\end{figure*}

\PP{Adversarial Visual Component Manipulation Attacks} 
To evade visual similarity-based phishing detectors, attackers manipulate visual components (\eg, logos)~\cite{Apruzzese2023SaTML, Yuan2024WWW, lee2023attacking}.
Particularly, Giovanni \etal~\cite{Apruzzese2023SaTML} found that phishers also bypass detectors by simply altering company name styles and stretching logos. Moreover, Ying \etal and Hao \etal~\cite{Yuan2024WWW} applied perturbations to visual components, and user study results demonstrated that these adversarial phishing techniques pose threats to both users and machine learning-based phishing website detectors. Recently, Lee \etal~\cite{lee2023attacking} developed an adversarial learning framework using imperceptible perturbation vectors based on a trained Vision Transformer (ViT)~\cite{dosovitskiy2021ViT} and a Swin Transformer (Swin)~\cite{liu2021swin}. Hao \etal~\cite{Look2024Hao} attacked the logos by changing fonts and generating adversarial logos through diffusion.
\vspace{-5px}
\section{Problem Statement} \label{sec:motivation}
Little effort has been made to systematically evaluate existing visual similarity-based defense models in real-world settings.
Prior research~\cite{lee2023attacking,Apruzzese2023SaTML} predominantly focus on presenting novel attack models rather than systematically identifying and analyzing new inherent vulnerabilities of the models.
Meanwhile, prior evaluation studies~\cite{Lin2021phishpedia,Liu2022phishintention,Abdelnabi2020visualphishnet,Afroz2011PhishZoo} do not consider quantifying each influential factor.

To bridge this gap, we conduct a comprehensive evaluation of the effectiveness of visual similarity-based anti-phishing models using a large-scale real-world phishing dataset comprising 451k websites, 4,190 sampled phishing websites, and 2,500 benign samples (of the Tranco Top 1000 websites). Then, we examine how attackers manipulate visual elements and test model resilience against systematically modified logo images. Our study aims to (1) assess the effectiveness of current visual similarity-based models against real-world phishing attacks;
(2) identify the root causes of the models' failures in classifying phishing websites; and (3) investigate new phishing tactics that manipulate visual components like logo images to circumvent existing visual similarity-based models.

% To bridge this gap, we comprehensively evaluate prominent visual similarity-based anti-phishing models using a large-scale real-world phishing dataset comprising 451k websites, 4,190 sampled phishing websites, and 2,500 benign samples (of the Tranco Top 1000 websites~\cite{Tranco:online}). Moreover, we investigate strategies attackers use to manipulate the visual components. To evaluate the models' robustness against these tactics, we semi-manually manipulate logo images according to our predefined rules. By testing the models on this dataset, we aim to uncover their strengths and weaknesses in detecting adaptive adversaries. Our study comprehensively analyzes the real-world effectiveness and robustness of visual similarity-based phishing detection models, offering insights for developing more resilient defenses against evolving phishing tactics. Our study aims to (1) assess the effectiveness of current visual similarity-based models against real-world phishing attacks;
% (2) identify the root causes of the models' failures in classifying phishing websites; and (3) investigate new phishing tactics that manipulate visual components like logo images to circumvent existing visual similarity-based models.

\section{Evaluation Design} 
\label{sec:experiment_setting}

\subsection{Overview of Our Evaluation Methodology} 
\label{sec:experiment:overview}

\begin{table}[!t]
\caption{\textbf{Training and Testing Reference List Dataset.} Training datasets are used to re-train the models.}
\label{tab:dataset}
%\vspace{-5px}
\begin{subtable}{1\linewidth}
    \setlength{\tabcolsep}{2pt}
    \centering
    \resizebox{1\textwidth}{!}{
    \begin{NiceTabular}{lllclll}
        \toprule
        \textbf{ }& \multicolumn{1}{c}{\textbf{Definition}} & \multicolumn{1}{c}{\textbf{Dataset Source}} & \multicolumn{1}{c}{\textbf{Target Model}} & \multicolumn{1}{c}{\textbf{\# Brand}} & \multicolumn{1}{c}{\textbf{\# Image}}\\ 
        \midrule
        \multirow{3}{*}{$\boldsymbol{R_{base}}$} & \multirow{3}{*}{Baseline Ref.}& PhishIntention~\cite{Liu2022phishintention} & \textbf{L-based} & 277 (\textbf{B}) & 3,064 (\textbf{B})\\
        & & PhishIntention~\cite{Liu2022phishintention} & \textbf{S-based} & 277 (\textbf{B}) & 9,530 (\textbf{B}) \\
        % Phishpedia & L & 181 (+) & 3,690 \\
        & &VisualPhishNet~\cite{Abdelnabi2020visualphishnet} & \textbf{S-based}  & 155 (\textbf{B}) 155 (\textbf{P}) & 9,363 (\textbf{B}) 1,193 (\textbf{P}) \\ 
        \midrule
        \multirow{2}{*}{$\boldsymbol{R_{ext}}$} & \multirow{2}{*}{Extended Ref.}&Extended Logo & \textbf{L-based} & 277 (\textbf{B}) & 3,167 (\textbf{B})\\
        & & Extended Screenshot & \textbf{S-based} & 277 (\textbf{B}) 213 (\textbf{P}) & 9,633 (\textbf{B}) 1,179 (\textbf{P})\\
        \midrule
        \multicolumn{6}{l}{\makecell[l]{
        *\textbf{L-based} = Logo-based models; \textbf{S-based} = Screenshot-based models;
        \textbf{\textbf{B}} = Benign; \textbf{\textbf{P}} = Phishing.\\
        }}
        \end{NiceTabular}
    }
        \caption{\textbf{Training Dataset}: Phishing Target Brand Reference Lists.}
        \label{tab:targetlist_info}
    \end{subtable}
% \end{table}
    \vspace{-5px}
\begin{subtable}{1\linewidth}
    \setlength{\tabcolsep}{6pt}
    \resizebox{1\columnwidth}{!}{%
    \begin{NiceTabular}{l l rrrr}
    \toprule
     \textbf{}&\multicolumn{1}{c}{\textbf{Type}} & \multicolumn{1}{c}{\textbf{\# Sample}} & \multicolumn{1}{c}{\textbf{\# Domain}} & \multicolumn{1}{c}{\textbf{\# Brand}} & \multicolumn{1}{c}{\textbf{\# Cluster}}\\
     \midrule
     \protect{$\boldsymbol{D_{learn}}$}& Only-Learned-Brand Dataset & 312,355 & 104,813 & 110 & 2,797  \\
     \protect{$\boldsymbol{D_{all}}$}& All-Brand Dataset & 451,514 & 163,864 & 270 & 4,190 \\
     \protect{$\boldsymbol{D_{sample}}$}& Sampled Dataset & 4,190 & 3,455 & 270 & 4,190 \\
     \protect{$\boldsymbol{D_{benign}}$}& Benign  Dataset & 2,500 & 100 & --- & --- \\
    \bottomrule
    \end{NiceTabular}
    }
        \caption{\textbf{Testing Dataset}: Collected Phishing and Benign Websites.}
    \label{tab:apwg_dataset}
    \end{subtable}
    \vspace{-10px}
\end{table}

We illustrate the overview of experiments in \autoref{fig:overview}. Our methodology includes the following key steps. \textit{First}, we develop a web crawler that collects the screenshots and client-side resources of real-world phishing websites using phishing URLs reported by the Anti-Phishing Working Group (APWG) \texttt{eCX}~\cite{APWGTheA96:online} (\BC{1}). 
Note that APWG \texttt{eCX} shares real-time phishing threat intelligence (\eg, phishing URLs).
Then, we refine the collected phishing dataset by removing unnecessary data (\eg, error pages) via clustering screenshots, as described in~\autoref{sec:dataset:collection}. 
Moreover, for the false positive evaluation, we also collect benign website samples based on the Tranco Top 1000 websites.
\looseness=-1

\textit{Second}, we standardize brand knowledge of the models using the same phishing target brand reference datasets (\BC{2}) to ensure fairness, which is $\boldsymbol{R_{base}}$, the baseline phishing target brand reference dataset from public sources. Considering the evolution of brands (\eg, rebranding logos or website layouts), we further expand it and yield $\boldsymbol{R}_{ext}$ our extended reference dataset (\BC{3}). \autoref{tab:targetlist_info} shows the statistics of these datasets. 
 
\textit{Third}, we carefully select seven popular visual similarity-based phishing defense models considering different factors, as detailed in~\autoref{sec:selection:models}.
Then, these datasets ($\boldsymbol{R}_{base}$ and $\boldsymbol{R}_{ext}$) are used to re-train the models (\BC{4}). Further details can be found in \autoref{sec:retraining:models}.
\textit{Fourth}, we evaluate the effectiveness of the seven models using our collected real-world datasets (\BC{5}): the ``Only-Learned-Brand'' dataset ($\boldsymbol{D_{learn}}$), the ``All-Brand'' dataset ($\boldsymbol{D_{all}}$), the ``Sampled Dataset'' ($\boldsymbol{D_{sample}}$), and the ``Benign Dataset'' ($\boldsymbol{D_{benign}}$) as shown in~\autoref{tab:apwg_dataset}.

Regarding the failed samples for the models (\eg, changing the text logo to upper case, as shown in~\autoref{fig:second_image}), we attempt to understand why the models fail to classify specific phishing attacks. We address RQ2 by designing another experiment where we manipulate visual components (\eg, logo images) of phishing websites in two ways (\BC{6}): (1) visible manipulation techniques (\ie, changing logo color or location) and (2) perturbation-based adversarial manipulation techniques (\ie, white-box attack). \autoref{tab:manipulation-vicomp-w-img} illustrates the examples of the manipulations. Then, we evaluate the robustness of models using the manipulated dataset and quantify their failures (\BC{7}), discussed in \autoref{sec:design:manipulating}.

\subsection{Real-world Phishing Dataset Collection}\label{sec:dataset:collection}
\PP{Web Crawler Design}
APWG \texttt{eCX}~\cite{APWGTheA96:online} is one of the most trusted, largest repositories for real-world phishing attacks as it aggregates reports from security vendors, financial institutions, and ISPs. 
It is widely used to analyze and better understand phishing ecosystems~\cite{kim2021security,oest2020sunrise,oest2020phishtime,zhang2021crawlphish,lim2024phishing}. 
As APWG \texttt{eCX} provides only phishing URLs, we newly design a web crawler that regularly (every minute) gathers (1) client-side resources (\eg, logos, HTML, etc.) of phishing websites and captures (2) the screenshots. 
The web crawler is implemented through \texttt{Google Selenium Chrome WebDriver}~\cite{Selenium} to simulate real user interactions with phishing websites, fully loading and rendering all client-side resources on the webpages. 
Additionally, \texttt{Selenium Chrome WebDriver} may assist in evading basic anti-bot techniques employed by phishing websites~\cite{amin2020web,li2021good}.

\PP{Refining Dataset}
APWG \texttt{eCX} provides a total of 15,747,193 (15.7M) real-world phishing URLs from July 2021 to July 2023 (25 months). 
Our crawler successfully accesses 6,118,654 (6.1M) phishing websites.  
61.1\% of inaccessible websites are due to server shutdowns or network errors such as DNS resolving errors.
Among 6.1M samples, we exclude internal error web pages (\ie, page not found) and improve label accuracy by clustering screenshots based on \texttt{Fastdup}~\cite{Fastdup}, an open-source tool that is effective in identifying duplicates, outliers, and clusters of related images by calculating the edge distances inside the graph component.
Specifically, we select 6,885 clusters with more than 20 screenshots, as these account for over 90\% of the total 6.1M screenshots. Two security researchers independently conduct manual inspections of the clusters. They each select three representative samples from each cluster and label them. The researchers then compare their results, discuss any discrepancies, and combine the clusters. This process is iteratively repeated until a consensus on all labels is reached. The filtered dataset contains 2,160,933 samples, representing 270 brands and 4,190 clusters.
\looseness=-1

\PP{Final Phishing Dataset for Evaluation} 
Among the filtered dataset, 
a small percentage (20\%) of clusters (\ie, merged brands) hold the majority (80\%) of total screenshots, which makes the evaluation process time-consuming and susceptible to bias. Therefore, we randomly select 1,000 samples for each cluster (if lower than 1,000, then all of them are selected) to ensure fairness. This process results in a total of 451,514 samples with 270 brands and 4,190 clusters, denoted as \protect{$\boldsymbol{D_{all}}$}.
Furthermore,
\protect{$\boldsymbol{D_{all}}$} includes some brands that are not present in the training dataset ($\boldsymbol{R_{base}}$ and $\boldsymbol{R_{ext}}$), meaning the brands are not learned by models. 
We identify 110 common brands between $\boldsymbol{D_{base}}$ and $\boldsymbol{D_{all}}$. This dataset, called $\boldsymbol{D_{learn}}$, includes 312K samples with 110 learned brands. 
Additionally, we sample 1 example for each cluster and construct $\boldsymbol{D_{sample}}$ to test \texttt{DynaPhish} due to computational intensity and extensive Google Search API costs.
In summary, $\boldsymbol{D_{learn}}$, $\boldsymbol{D_{all}}$, $\boldsymbol{D_{sample}}$, and $\boldsymbol{D_{benign}}$ are used to better understand the models’ performance (\ie, effectiveness and robustness) in real-world scenarios and to examine the impact of data on unlearned brands.
\looseness=-1

\PP{Benign Website Dataset for False Positive Evaluation.}
To evaluate false positive rates of phishing detection models, we assemble a dataset of legitimate websites. We randomly select 100 domains from the top 1,000 websites in the Tranco 1M ranking. For each domain, we collect monthly snapshots between July 2021 and July 2023 using the Internet Archive's Wayback Machine (\url{https://archive.org/}), capturing URLs, screenshots, and HTML content. This process yields 2,500 benign samples (100 domains $\times$ 25 months). This dataset includes 14 domains comprising 350 samples, which are associated with 12 target brands present in the training data ($\boldsymbol{D_{learn}}$).

\subsection{Model Selection for Evaluation}
\label{sec:selection:models}
We carefully select representative models of visual similarity-based anti-phishing techniques for comprehensive evaluations. 
\textit{First}, we initially search some keywords (\ie, `anti-phishing,' `phishing detection,' and `visual-based similarity') at the top-tier security, computer vision, and machine learning conferences to identify model candidates.
Model candidates are summarized in \autoref{tab:referece_based_model} and \aref{subsec:model_summary}.
% \autoref{tab:referece_based_model}.
\looseness=-1

From these candidates, we choose models with publicly available code to ensure fidelity to the original papers. 
% Furthermore, we also consider citations and the time and money cost of testing the large-scale dataset. 
Furthermore, considering the utilized information and pipeline structures, we select three recent popular logo-based phishing detection models, \texttt{DynaPhish}~\cite{Dynaphish2023Liu}, \texttt{PhishIntention}~\cite{Liu2022phishintention}, and \texttt{Phishpedia}~\cite{Lin2021phishpedia}. They use URLs, screenshots, and HTML as input, employing cropped logos to match the logos in the reference list for brand detection and identification. Additionally, \texttt{PhishZoo}~\cite{Afroz2011PhishZoo} uses the same inputs but matches between screenshots and logos.
For a broader comparison, we select \texttt{Involution}~\cite{viniavskyi2022openglue}, a model not specifically tailored for phishing detection. We are also interested in whole image comparison and thus select
\texttt{VisualPhishNet}~\cite{Abdelnabi2020visualphishnet}, which detects phishing using screenshots, and \texttt{EMD} (Earth Mover’s Distance)~\cite{Fu2006EMD} which is a static model for screenshot-based phishing detection.
Finally, detailed explanations of the models, including their descriptions and inputs, are provided in~\aref{subsec:model_summary} and~\aref{sec:select_model}.

% Specifically, our selection includes three popular logo-based phishing detection models, \texttt{DynaPhish}~\cite{Dynaphish2023Liu}, \texttt{PhishIntention}~\cite{Liu2022phishintention}, and \texttt{Phishpedia}~\cite{Lin2021phishpedia}. Additionally, we include \texttt{VisualPhishNet}~\cite{Abdelnabi2020visualphishnet}, which detects phishing using screenshots. To provide a comprehensive comparison, we incorporate two approaches with high citations, \texttt{EMD} (Earth Mover’s Distance)~\cite{Fu2006EMD} for screenshot-based phishing detection and \texttt{PhishZoo}~\cite{Afroz2011PhishZoo} for logo-based detection. Finally, we include \texttt{Involution}~\cite{viniavskyi2022openglue}, a model not specifically tailored for phishing detection, to provide a broader baseline for comparison. Detailed model information can refer to~\aref{subsec:model_summary} and~\aref{sec:select_model}.

\subsection{Re-training Models \& Evaluation Plan}
\label{sec:retraining:models}
We aim to rigorously evaluate the effectiveness of the carefully selected seven visual similarity-based anti-phishing models with our extensive dataset of real-world phishing websites. Initially, training under varying conditions and with diverse reference lists can significantly impact the evaluation outcomes. Additionally, based on our evaluation, the presence of either outdated or new visual elements, such as rebranded logo images or login forms, can profoundly affect model performance, as these elements might not have been adequately captured during initial training. 
For example, updates (\ie, refreshes) to Facebook's login form, user interface, or icons could potentially adversely impact the model's performance. To ensure a more equitable and cautious approach in evaluations and model performance comparisons, we \textit{re-train} the models by incorporating them with the same reference knowledge of brands, taking these factors into account.

\PP{Two Variants of Re-trained Models}
Our objective in developing two variants of re-trained models is to assess the impact of brand variations (\eg, refreshed or outdated logos) because our initial evaluation of the all-brands dataset with the base reference list demonstrates the limitations of the outdated reference list (\ie, refreshed or outdated logos are not included). 
The distinction between the two variants lies in the reference lists used: (1) the baseline phishing target brand reference dataset ($\boldsymbol{R_{base}}$) and (2) our expanded reference dataset ($\boldsymbol{R_{ext}}$), as detailed in~\autoref{tab:targetlist_info}. 
$\boldsymbol{R_{base}}$ is the same brand list as \texttt{PhishIntention} (L-based) for logo-based anti-phishing models and \texttt{PhishIntention} (S-based) for screenshot-based models. $\boldsymbol{R_{ext}}$ is obtained by expanding the $\boldsymbol{R_{base}}$ by adding a newly updated logo and screenshot variance from Archive between 2016 and 2023.
Note that we train \texttt{VisualPhishNet} only on the $\boldsymbol{R_{ext}}$ screenshot dataset, as it requires both benign and phishing data during the training phase and ensures the brand knowledge consistent. During evaluation, we integrate the trained model with \texttt{PhishIntention} (S-based) and $\boldsymbol{R_{ext}}$ screenshot dataset as the reference lists for baseline and extended results, respectively.

\subsection{Manipulating Visual Component (Logo)}
\label{sec:design:manipulating}

Through our evaluation experiment, we analyze the manipulation tactics employed by phishing attackers. 
We find that there are four primary components typically exploited by adversaries in phishing attacks, including logos, pop-ups, login forms, and others, as presented in~\autoref{tab:pass_screenshot_category} of~\aref{sec:pass_example_categorization}.
Upon randomly selecting images from the failure results, we discern that logos are prevalent targets used by adversaries to circumvent detection mechanisms. Furthermore, logos serve as the indicators for both users and detection mechanisms to recognize the websites. 
Consequently, this study focuses primarily on the logo component. 

\subsubsection{Manipulation Methods}
\label{subsubsec:coarse_sample}

Phishing attackers not only aim to mimic legitimate target brand websites to deceive potential victims closely but also aim to evade detection by anti-phishing systems, particularly those based on visual similarity. To achieve this, they have developed various adversarial visual component manipulation strategies. We broadly categorize such strategies into (1) visible manipulation techniques and (2) perturbation-based adversarial manipulation techniques.

\PP{Visible Manipulation Methods} 
The visible manipulation techniques involve substantial, noticeable changes to the original visual appearance, such as altering the image color~\cite{ke2023neural}, text~\cite{yang2020swaptext}, and UI design patterns~\cite{Subramani2022PhishInPatterns}. 
For instance, logos (\eg, changing the Facebook logo's font and converting the letters to uppercase) are manipulated, as illustrated in~\autoref{fig:facebook_attack_example}.
Based on the failure samples discussed in \autoref{sec:results}, we identify 13 types of manipulations used by adversaries in real datasets and choose SRNet~\cite{WuEdit2019} as an additional deep learning-based method in this category. For descriptions of SRNet, please refer to~\aref{sec:perturbated_example}. Subsequently, we craft logos corresponding to these manipulations and attach them to the original screenshots. 
Specifically, we use `remove.bg' (\url{https://www.remove.bg/}) to eliminate the background of logos.
If users are not looking at the logos carefully, they readily overlook it and can be readily lured.
The manipulations and crafted logos are shown in~\autoref{tab:manipulation-vicomp-w-img}. 
Note that we do not combine more than two visible manipulation ways.
% \looseness=-1

\PP{Perturbation-based Manipulation Methods} 
Perturbation-based adversarial manipulation techniques introduce subtle manipulations~\cite{szegedy2013intriguing, Goodfellow2015FGSM, madry2018towards, Carlini2017CW} that are difficult for humans to detect visually.
We introduce perturbations to the logos using popular white-box and black-box attack techniques.
The perturbed logos are then returned to their original positions on the screenshots, ensuring a seamless integration into the visual context. For descriptions of selected models and perturbated logos, please refer to~\aref{sec:perturbated_example} and~\autoref{fig:craft_perturbated}. 
\looseness=-1

\subsubsection{Evaluation Plan for Manipulation}
All models used to evaluate robustness are trained on the `Extended Ref.' ($\boldsymbol{R_{ext}}$) or `Baseline Ref.' ($\boldsymbol{R_{base}}$) datasets along with the original needed datasets. 
We focus on the learned 110 brands ($\boldsymbol{D_{learn}}$) to obtain more accurate results. 
Furthermore, to investigate the impact of different factors (URLs, Logo, and HTML), we conduct an ablation study for four models that highly depend on the three factors. Specifically, we manually collect the URLs, HTML, and screenshots of the login page or main page of 110 brands' websites. Then, we use \texttt{typosquatter} (\url{https://github.com/typosquatter/ail-typo-squatting}) to generate various domain typos to replace their original domain within URLs.
Finally, the total dataset contains 6,569 visible manipulated screenshots, 544 perturbated screenshots, 110 benign URLs, and 1,321 squatted URLs for 110 brands.
These URLs are then paired with corresponding altered images to curate the phishing testing dataset.
\looseness=-1

\section{Evaluation of Model Performance on Real-world Phishing Datasets} \label{sec:results}

\begin{table*}[!t]
    \setlength{\tabcolsep}{7pt}
    \renewcommand{\arraystretch}{1.05}
    \caption{
   \textbf{ Phishing Detection Results} on 312,355 ($\boldsymbol{D_{learn}}$) and 451,514 ($\boldsymbol{D_{all}}$) testing samples from APWG. \textbf{Phishing Brand Identification Results} on 312,355 ($\boldsymbol{D_{learn}}$). The bold numbers denote the better detection or identification rate.}
    \label{tab:result_phishing_detection}
    %\vspace{-5px}
    \centering
    \resizebox{1\linewidth}{!}{
    \begin{NiceTabular}{lrrrrrrrrrc}
        \toprule
        \multirow{3}{*}{\textbf{Model}}  & \multicolumn{4}{c}{\textbf{Phishing Detection}} & \multicolumn{6}{c}{\textbf{Phishing Brand Identification (with $\boldsymbol{D_{learn}}$)}}\\ 
        \cmidrule(lr){2-5} \cmidrule(lr){6-11}
         & \multicolumn{2}{c}{\textbf{Only-Learned Brands} \textbf{(312,355)} $\boldsymbol{D_{learn}}$} & \multicolumn{2}{c}{\textbf{All Brands (451,514) $\boldsymbol{D_{all}}$}} & \multicolumn{3}{c}{\textbf{Baseline Ref.} ($\boldsymbol{R_{base}}$)} & \multicolumn{3}{c}{\textbf{Ext. Ref.} ( $\boldsymbol{R_{ext}}$)}\\
        \cmidrule(lr){2-3}\cmidrule(lr){4-5}\cmidrule(lr){6-8}\cmidrule(lr){9-11}
         & \multicolumn{1}{c}{\textbf{Baseline Ref.} ($\boldsymbol{R_{base}}$)} & \multicolumn{1}{c}{\textbf{Ext. Ref.} ($\boldsymbol{R_{ext}}$)} & \multicolumn{1}{c}{\textbf{Baseline Ref.} ($\boldsymbol{R_{base}}$)} & \multicolumn{1}{c}{\textbf{Ext. Ref.} ($\boldsymbol{R_{ext}}$)} & \multicolumn{1}{c}{$\boldsymbol{I_{tp}}$$^1$} & \multicolumn{1}{c}{$\boldsymbol{I_{tp} / N_p}$$^2$} & \multicolumn{1}{c}{$\boldsymbol{I_{tp} / N_{tp}}$$^3$} & \multicolumn{1}{c}{$\boldsymbol{I_{tp}}$$^1$} & \multicolumn{1}{c}{$\boldsymbol{I_{tp} / N_p}$$^2$} & \multicolumn{1}{c}{$\boldsymbol{I_{tp} / N_{tp}}$$^3$}\\ 
        \midrule
        % \multirow{5}{*}{\makecell[l]{Total \\451,514}} 
        PhishIntention&      204,880 \xspace(65.59\%) & 206,846 (66.22\%) &  235,838 (52.23\%) & 237,861 (52.68\%) & 200,134 & 64.07\% & 97.68\% & 202,123 & 64.71\% & 97.72\%\\ % ~\cite{Liu2022phishintention}
        Phishpedia& 232,572 \xspace (74.46\%) & \textbf{274,779 (87.97\%)} & 275,292 (60.97\%) & 318,196 (70.47\%) & 222,860 & 71.34\% & 95.82\%& \textbf{265,627} & \textbf{85.04\%} & 96.67\%\\ % ~\cite{Lin2021phishpedia}      
        Involution&  \textbf{253,965 (81.31\%)} & 264,782 (84.77\%)& 289,058 (64.02\%) & 301,035 (66.67\%)  & \textbf{253,090} & \textbf{81.03\%} & \textbf{99.66\%} &263,835 & 84.47\% & \textbf{99.64\%}\\ % ~\cite{Li2021CVPR}  
        PhishZoo & 241,206 \xspace(77.22\%) & 269,748 (86.36\%) & \textbf{353,292 (78.25\%)} & \textbf{389,585 (86.28\%)} &  30,829 & 9.86\% & 12.78\% & 89,724 & 28.73\% & 33.26\%\\ % ~\cite{Afroz2011PhishZoo}       
        VisualPhishNet & 122,106\xspace (39.09\%) & 126,762 (40.58\%) & 181,177 (40.13\%) & 186,606 (41.33\%) &81,119 & 25.97\% & 66.43\% & 83,697 & 26.80\% & 66.03\% \\ % ~\cite{Abdelnabi2020visualphishnet}
        EMD & 95,632 \xspace (30.62\%) & 97,880 (31.34\%) & 133,241 (29.51\%) & 136,697 (30.28\%) & 22,478 & 7.20\% & 23.50\% & 22,426 & 7.18\% & 22.91\% \\ % ~\cite{Fu2006EMD}
        \midrule
        \multicolumn{11}{l}{\makecell[l]{
        $^1$\textbf{$I_{tp}$} = The number of phishing samples brands correctly identified;$\>\>$
        $^2$\textbf{$I_{tp} / N_p$} = The phishing target brand identification rate out of the total phishing testing samples; \\
        $^3$\textbf{$I_{tp} / N_{tp}$} = The phishing target brand identification rate out of the only samples detected as phishing by each model.}}
        \end{NiceTabular}
        }
        %\vspace{-10px}
\end{table*}

\begin{table}[!h]
    \renewcommand{\arraystretch}{1}
     \normalsize
    \caption{
    \textbf{Phishing Detection and Identification Results on  $\boldsymbol{D_{sample}} (4,190)$ from APWG trained with $\boldsymbol{R_{base}}$.}}
    \label{tab:dynaphish_apwg}
    \centering
    \resizebox{1\linewidth}{!}{
    \begin{NiceTabular}{lrr}
        \toprule
        \multicolumn{1}{c}{\textbf{Model}}  & \multicolumn{1}{c}{\textbf{Detection}} & \multicolumn{1}{c}{\textbf{Identification $\boldsymbol{I_{tp}}$ ($\boldsymbol{I_{tp} / N_{tp}}$)}}\\ 
        \midrule
        PhishIntention  &   2,056 \xspace (49.07\%)& \textbf{2,027\xspace(98.56\%)}\\
        Phishpedia  &   2,395 \xspace (57.16\%) & 2,212\xspace(92.36\%)\\
        Involution  &   2,538 \xspace (60.57\%)& 2,470\xspace(97.32\%)\\
        PhishZoo    &   \textbf{3,190 \xspace (76.13\%)}&306\xspace(9.59\%)   \\
        VisualPhishNet  &   1,418 \xspace (33.84\%)&773\xspace(54.51\%)  \\
        EMD &   1,150 \xspace (27.45\%)&236\xspace(20.42\%)  \\
        DynaPhish   &   923 \xspace (22.03\%)& 904\xspace(97.94\%)   \\
        \bottomrule
        % \multicolumn{3}{l}{\makecell[l]{
        % $^1$\textbf{$I_{tp}$} = The number of brands of phishing samples correctly identified;$\>\>$\\
        % $^2$\textbf{$I_{tp} / N_{tp}$} = The phishing target brand identification rate out of the total \\phishing testing samples.}}
        \end{NiceTabular}
        }
        % \vspace{-10px}
\end{table}

We first assess the effectiveness of seven models using our real-world phishing and benign datasets in phishing detection and computational costs using FLOPs and parameters (\autoref{subsubsec:apwg_results}), false positive rates (\autoref{sec:false:positive}), and phishing brand identification (\autoref{sec:brand:identification:result}). 
Additionally, we further analyze the reasons for the failures of the models (\autoref{sec:In-depth Analysis of Detection Failures}). Finally, we conduct ablation studies on logo, URL, and HTML features to understand their contributions to the detection performance (\autoref{sec:Ablation_Study}).

\PP{Settings.}
$\boldsymbol{D_{learn}}$, $\boldsymbol{D_{all}}$,  $\boldsymbol{D_{sample}}$, and $\boldsymbol{D_{benign}}$
in~\autoref{tab:apwg_dataset} are leveraged to evaluate the performance of the models.
We define a \textit{true positive} as correctly detecting phishing and a \textit{false positive} as incorrectly classifying a benign website as phishing. For phishing detection, let $N_p$ be the number of phishing testing samples, $N_{tp}$ (\textit{tp} stands for true positive) be the number of correctly classified phishing samples, and $I_{tp}$ be the number of correctly identified target brands. For benign samples, let $N_b$ be the total number, $N_{fp}$ (\textit{fp} stands for false positive) be the number falsely classified as phishing, and $I_{fp}$ be the number with incorrectly identified target brands.

Using these metrics, we calculate six rates: (1) the true positive rate ($N_{tp}/N_p$), measuring phishing detection accuracy; (2) the phishing identification rate ($I_{tp}/N_p$) and (3) identified phishing accuracy ($I_{tp}/N_{tp}$) for brand identification; (4) the false positive rate ($N_{fp}/N_b$), measuring benign misclassification; (5) the false identification rate ($I_{fp}/N_{fp}$) and (6) the overall false brand rate ($I_{fp}/N_b$) for incorrect brand identification.
\looseness=-1

\PP{Thresholds} 
Threshold values are obtained by prior work~\cite{lindsey932020online} and our further check with their datasets. 
Specifically, we set the thresholds as 0.83, 0.83, 0.83, 40, 1, 0.94, and 0.7 for \texttt{DynaPhish}, \texttt{PhishIntenion}, \texttt{Phishpedia}, \texttt{PhishZoo}, \texttt{VisualPhishNet}, \texttt{EMD}, and \texttt{Involution}, respectively, to identify potential phishing instances effectively.

\subsection{Result: Detection Effectiveness}
\label{subsubsec:apwg_results} 
Phishing detection refers to the capability to correctly classify websites as phishing or legitimate. 
\autoref{tab:result_phishing_detection} shows the phishing detection and brand identification results on large-scale real-world datasets ($D_{learn}$ and $D_{all}$). 
\autoref{tab:dynaphish_apwg} provides the phishing detection and brand identification results of $D_{sample}$.
\autoref{tab:achive_phishing_detection} describes the false positive and cost results.

\PP{General Performance}
All seven models demonstrate lower performance in phishing detection, compared to their originally reported results. Specifically, when trained on $\boldsymbol{R}_{base}$, six models in~\autoref{tab:result_phishing_detection} failed to detect 38.62\% of the 312,355 phishing samples from their learned target brands ($\boldsymbol{R_{learn}}$). Even with expanded training on $\boldsymbol{R}_{ext}$, these models still miss 33.8\% of phishing websites.

Logo-based phishing detection models experience significant performance degradation on comprehensive datasets. True positive rates of \texttt{PhishIntention}, \texttt{Phishpedia}, and \texttt{Involution} drop by 13--19\% when tested on the whole-brand dataset ($D_{all}$, $R_{ext}$).
Although \texttt{PhishZoo} reaches 77.22\% accuracy on limited datasets ($D_{learn}$) and 78.25\% on $D_{all}$, the false positive reaches 93.92\% on $D_{benign}$. Their primary weakness stems from poor resilience to logo variations and heavy reliance on static reference lists for similarity matching, making them vulnerable to evasion through logo modifications not present in their reference lists.

\observ{
Reference list-based models can introduce weaknesses. Logos or screenshots not included in the reference list but known to users may mislead the detection models. This highlights the necessity of expanding and continuously updating the reference lists and detection models}

\PP{Learned Vs. Unlearned Testing Dataset}  
When deployed in real-world environments, phishing detection models are highly likely to encounter unlearned brands.
To further investigate the models' readiness for real-world deployment with new, unknown phishing websites, we compare the results between \boldsymbol{$D_{learn}$} (containing only learned brands) and  \boldsymbol{$D_{all}$} (also including unlearned, new brands).

The results, detailed in \autoref{tab:result_phishing_detection}, 
show a decline in detection performances in more challenging scenarios (\boldsymbol{$D_{all}$}).
Particularly, \texttt{PhishIntention}, \texttt{Phishpedia}, and \texttt{Involution} (logo-based models) decrease in the detection rate from 66.22\% to 52.68\% (13.54\%{\small\faLongArrowDown}),  from 87.97\% to 70.47\% (17.5\%{\small\faLongArrowDown}), and from 84.77\% to 66.67\% (18.1\%{\small\faLongArrowDown}), respectively with \boldsymbol{{$R_{ext}$}}.
The three models rely on identifying and comparing logo similarities with their target brand reference list.
If either these models fail to recognize the logo or the brand does not appear in the reference list, the similarity score will be lower than the threshold, leading to detection failures.

\texttt{PhishZoo} shows consistent but unreliable performance across datasets, with low identification rates (9.86\% on $D_{learn}$ with $R_{base}$, 9.59\% on $D_{sample}$) and high false positives (93.92\% on $D_{benign}$). This indicates misclassification rather than accurate detection, stemming from its HTML/URL keyword selection and SIFT feature comparison methodology.
Screenshot-based models (\texttt{EMD}, \texttt{VisualPhishNet}) demonstrate better resilience to unlearned brands compared to logo-based approaches. \texttt{EMD}'s detection rate slightly decreases from 31.34\% to 30.28\% (1.06\%{\small\faLongArrowDown}), while \texttt{VisualPhishNet} shows a marginal increase from 40.58\% to 41.33\% (0.75\%{\small\faLongArrowUp}) on $R_{ext}$. However, these models generally achieve lower detection rates than logo-based detectors (\texttt{PhishIntention}, \texttt{Phishpedia}, \texttt{Involution}), highlighting a trade-off between resilience to unlearned brands and overall detection performance.

\observ{Logo-based models heavily rely on their pre-established brand reference lists, leading to degraded detection performance when encountering unlearned brands. In contrast, screenshot-based models demonstrate better resilience to unlearned brands, though they generally achieve lower detection rates than logo-based models}

\PP{Baseline Vs. Extended Reference List} 
Recall that the Extended Reference List Dataset (\boldsymbol{$R_{ext}$}) is curated by manually adding more logo variance and screenshots of their learned target brands to the baseline dataset (\boldsymbol{$R_{base}$}).
Typically, the new logos are slightly changed from their prior logos.

We observe a significant performance increase in both \texttt{Phishpedia} and \texttt{PhishZoo} when being tested on \boldsymbol{$D_{learn}$}.
Specifically, the phishing detection accuracy of \texttt{Phishpedia} and \texttt{PhishZoo} increased from 74.46\% to 87.97\% (13.51\%{\small\faLongArrowUp}) and from 77.22\% to 86.36\% (9.14\%{\small\faLongArrowUp}), respectively.
The performance gain for \texttt{Phishpedia} is attributed to the recent logo updates in the dataset, highlighting the importance of comprehensive and regularly updated logo collections in phishing detection model design.
In contrast, the apparent improvement in \texttt{PhishZoo} does not truly reflect its effectiveness in phishing detection. Considering its low identification results, it appears that logos among reference lists mislead \texttt{PhishZoo} to incorrectly recognize brands of phishing websites.
Moreover, \texttt{VisualPhishNet} and \texttt{EMD} demonstrate resilience to changes in their reference lists, as newly added screenshots often share visual styles with existing samples. However, these models face a fundamental limitation: the challenge of capturing the complete diversity of webpage layouts and designs, which constrains their overall performance.

These findings reveal a potential weakness in reference-based models that attackers could exploit using newly updated logos or various designs not included in the reference list.
While increasing reference data (regularly adding new logos) improves detection performance, it also prolongs computing time, presenting a crucial trade-off. 
This trade-off underscores the need for a strategic model design, where balancing detection efficacy and computational efficiency is vital.

\observ{Updating the reference list dataset to include varied logos, as well as retaining outdated logos, significantly improves the performance of logo-based models. However, expanding the reference dataset also increases computation time, necessitating a balance between detection accuracy and efficiency in model design}

\PP{Influence of Model Architecture} \texttt{PhishIntention}, \texttt{Phishpedia}, and \texttt{DynaPhish} share a common architectural foundation for phishing detection yet exhibit distinct performance characteristics. 
%
% While \texttt{Phishpedia} demonstrates superior detection rate with 57.16\% on the $D_{sample}$, both \texttt{DynaPhish} and \texttt{PhishIntention} achieve lower detection rate (22.03\% and 49.07\%). 
%
\texttt{Phishpedia} demonstrates a better detection rate (57.16\%) on the $D_{sample}$ compared to \texttt{DynaPhish} (22.03\%) and \texttt{PhishIntention} (49.07\%).
This performance divergence stems from the enhanced verification mechanisms implemented in \texttt{PhishIntention}, specifically their analysis of credential forms in HTML and screenshots. 
Notably, \texttt{DynaPhish} achieves the lowest detection rate (22.03\%) compared to architecturally similar models (\texttt{PhishIntention} and \texttt{Phishpedia}).
Our analysis reveals that 43.94\% of samples are flagged due to forbidden words appearing in the searched page titles. 
This filtering mechanism significantly impairs the model's overall detection performance.

\texttt{PhishZoo}, leveraging extracted keywords from URLs and HTML sources to mitigate false positives and employing the SIFT feature to calculate similarity in the target list, has reported promising results in phishing detection tasks.
However, the performance on brand identification reveals that the high accuracy initially indicated may be an overestimation of its true capabilities.
We reveal that the keyword selection approach based on TF-IDF scoring does not accurately capture brand-specific keywords that are highly indicative of phishing attempts. Specifically, we identify some examples where common words like `the' and `in' influence classification decisions, exposing a limitation in the feature engineering process.
\looseness=-1

Furthermore, the results indicate that screenshot-based reference methods are more stable for brand changes in both testing and reference data compared to logo-based methods, though their overall performance is worse. \texttt{EMD} measures the distribution distance between testing samples and the reference list, while \texttt{VisualPhishNet} utilizes a triplet Convolutional Neural Network to compare two screenshots. Unlike logos, the wide variety of screenshots presents a significant challenge in covering the full range of variations in the reference target list. Additionally, screenshots not present in the target list but sharing similar designs or features exhibit a high degree of resemblance to those in the target list. This highlights a potential flaw in screenshot-based methods due to the vast diversity and similarity among web designs.

\observ{
Inaccurate keyword extraction can degrade performance, while diverse screenshots and similar web designs present challenges to screenshot-based methods. Selecting an appropriate model structure is crucial to optimizing performance and mitigating these weaknesses}

\begin{table}[!t]
    \setlength{\tabcolsep}{7pt}
    \renewcommand{\arraystretch}{1.05}
    \caption{\textbf{False Positive Result and Performance Comparison.} Bold text indicates the best performance for each metric.}
    \label{tab:achive_phishing_detection}
    %\vspace{-5px}
    \centering
    \resizebox{1\linewidth}{!}{
    \begin{NiceTabular}{l|c|ccc}
        \toprule
        \multicolumn{1}{c}{\textbf{Model}}  & \multicolumn{1}{c}{$\boldsymbol{N_{fp} / N_b}$} & \multicolumn{1}{c}{\textbf{InferTime$^*$}} & \multicolumn{1}{c}{\textbf{FLOPs}} & \multicolumn{1}{c}{\textbf{Parameter}}\\ \midrule
        DynaPhish & \textbf{0} & 13.36s$^{**}$ & 215.66G & 88.92M\\
        PhishIntention&  \textbf{0} & 0.24s  & 215.66G & 88.92M\\
        Phishpedia &  406 (16.24\%) & 0.34s & 212.35G & 65.40M\\
        Involution &  99 (3.96\%) & \textbf{0.1s} & 212.67G & 53.04M\\
        PhishZoo$^{***}$ & 2,348 (93.92\%) & 16 mins & --- & ---\\
        VisualPhishNet & 338 (13.52\%) & 2.27s & \textbf{92.49G} & \textbf{21.27M}\\
        EMD$^{***}$ &  659 (26.36\%) & 23.54s & --- & ---\\
        \bottomrule
        \multicolumn{5}{l}{$^*$\textbf{InferTime}: Inference time taken for each sample.} \\
        \multicolumn{5}{l}{$^{**}$This time includes  online search via Google Search APIs.} \\
        \multicolumn{5}{l}{$^{***}$Note that FLOPs and parameters are unavailable for PhishZoo and EMD.}
        % \multicolumn{5}{l}{
        % Internet-dependent operations introduce uncontrollable variability.} \\
        \end{NiceTabular}
        }
        % \vspace{-10px}
\end{table}

\PP{Costs for Models} 
We evaluate model efficiency through inference time, FLOPs, and parameter size using a dataset of 2,500 benign samples ($D_{benign}$), with results presented in \autoref{tab:achive_phishing_detection}; detailed performance metrics for key components are provided in \autoref{tab:flop_parameter} and \aref{sec:flop_detail}. 
%Note that \texttt{Phishpedia} serves as the foundational architecture for logo detection in both \texttt{PhishIntention}, \texttt{DynaPhish}, and \texttt{Involution}. As \texttt{DynaPhish} directly adopts \texttt{PhishIntention} for the offline logo detection component without any modifications, both models share the same FLOPs and parameters.

\texttt{Involution} demonstrates the
% superior performance with the 
fastest inference time (0.1 seconds/sample), utilizing 212.67G FLOPs and 53.04M parameters. While \texttt{PhishIntention} and \texttt{DynaPhish} share identical architecture and computational requirements, \texttt{DynaPhish}'s additional Google Search verification process results in 56 times longer inference time.

Despite \texttt{Phishpedia}'s lower computational requirements compared to \texttt{PhishIntention}, its inference time is slower (0.34 seconds/sample) due to 97\% of samples bypassing CRP detection. \texttt{PhishZoo} shows the longest inference time (16 minutes/sample) due to sequential keyword comparisons and SIFT feature extraction. Though \texttt{VisualPhishNet} uses fewer computational resources than \texttt{Phishpedia}, its full-screenshot analysis approach leads to 6.68 times longer inference times compared to \texttt{Phishpedia}'s logo-focused analysis. Similarly, \texttt{EMD}'s screenshot-based approach results in significantly longer processing times.

\subsection{Result: False Positive Analysis}
\label{sec:false:positive}
% We evaluate the models' reliability through a comprehensive false positive analysis, a critical aspect for real-world deployment. 
Even models achieving high true positive rates may prove impractical if they generate excessive false alarms (\ie, incorrectly flagging benign websites as phishing). Such misclassifications can severely impact system efficacy in production environments, potentially undermining user trust and increasing operational overhead.

Our analysis in~\autoref{tab:achive_phishing_detection} reveals significant performance variations across models. 
\texttt{PhishZoo} demonstrates poor reliability with a 93.92\% false positive rate due to screenshot-logo mismatches during brand identification. While \texttt{DynaPhish} and \texttt{PhishIntention} achieve 0\% false positive rates, consistent with their original papers.
Their superior performance is due to additional verification steps but at the cost of true positives, 22.03\% and 49.07\% respectively, as shown in \autoref{tab:dynaphish_apwg}. Their performance is notably influenced by dataset characteristics, where $D_{sample}$ contains 2,155 samples that lack CRPs and 1,841 samples fall within forbidden domains for \texttt{DynaPhish}.
\texttt{Phishpedia} shows moderate performance with a 16.24\% false positive rate and 57.16\% phishing detection accuracy. Its limited knowledge scope, covering only 14 domains across 12 brands in $D_{benign}$, leads to poor generalization and frequent misclassification of legitimate websites from unfamiliar brands, significantly impacting its real-world applicability.

\observ{Detection models face trade-offs between false positives and detection accuracy. Importantly, most models mistakenly recognize benign websites outside knowledge scope as existing brands, highlighting significant deployment challenges in open set recognition}

%%%%---------------

\subsection{Result: Phishing Brand Identification} 
\label{sec:brand:identification:result}
Phishing brand identification refers to identifying brands that phishing websites attempt to impersonate. 
\autoref{tab:result_phishing_detection} shows the results on the learned brand dataset \boldsymbol{$D_{learn}$} and \autoref{tab:dynaphish_apwg} contains the results on \boldsymbol{$D_{sample}$}. 

\PP{General Performance}
Our analysis reveals significant differences in performance between logo-based and screenshot-based approaches for brand identification. The logo-based models (\texttt{Phishpedia} and \texttt{Involution}) demonstrate superior performance (84\%--88\% detection rate while 96\%--99.64\% on for identification on $D_{learn}$ with $R_{ext}$) across both tasks, highlighting the critical role that logo recognition plays in accurately identifying target brands and detecting phishing.

As shown in \autoref{tab:result_phishing_detection} and~\autoref{tab:dynaphish_apwg}, 
% logo-based models (\texttt{DynaPhish}, \texttt{PhishIntention}, \texttt{Phishpedia}, and \texttt{Involution}) demonstrate high target brand recognition rates among correctly classified phishing samples. 
with $R_{base}$ reference dataset on $D_{learn}$, \texttt{PhishIntention}, \texttt{Phishpedia}, and \texttt{Involution} achieve identification rates of 97.68\%, 95.82\%, and 99.66\% respectively. Similarly, on $D_{sample}$, they maintain strong performance with rates of 97.94\% (\texttt{DynaPhish}), 98.56\% (\texttt{PhishIntention}), 92.36\% (\texttt{Phishpedia}), and 97.32\% (\texttt{Involution}). This consistency across datasets indicates suitable reference logo coverage for recognized brands.

In contrast, screenshot-based models (\texttt{VisualPhishNet} and \texttt{EMD}) demonstrate inferior performance, with detection rates of 40.58\% and 31.34\%, and identification rates of 66.03\% and 22.91\% respectively for correctly detected samples on $D_{learn}$ with $R_{ext}$. This underperformance stems from their broader analysis of webpage elements rather than focus logo detection, complicated by the challenge of maintaining current screenshot datasets amid dynamic webpage layouts. However, these approaches maintain consistent effectiveness with unfamiliar brands where logos are unavailable, offering a valuable advantage despite lower overall performance.

\texttt{PhishZoo} achieves detection rates of 76.13\% on $D_{sample}$ and 86.36\% on $D_{learn}$ with $R_{ext}$, but its brand identification rate is only 9.59\% on $D_{sample}$. This disparity indicates misleading performance metrics, stemming from \texttt{PhishZoo}'s SIFT-based methodology that struggles with screenshot-to-logo matching in its brand database. In contrast, \texttt{VisualPhishNet} outperforms \texttt{EMD} in identification by employing triplet CNN to learn intra-brand similarities and inter-brand differences, while \texttt{EMD}'s effectiveness diminishes when brand screenshot distributions show high similarity.

\observ{Logo-based models currently offer the most reliable approach for standard phishing detection and brand identification, but they are susceptible to additional checking steps, used features, and logo components. Screenshot-based models struggle with web design diversity but may serve as a complementary solution for scenarios involving unknown or emerging brands}

\PP{Baseline Vs. Extended Reference List}
The six evaluated models show significant brand identification failure rates: approximately 51\% with Extended Reference dataset ($\boldsymbol{R_{ext}}$) and 57\% with Baseline Reference dataset ($\boldsymbol{R_{base}}$). Logo-based methods demonstrate improved performance with $\boldsymbol{R_{ext}}$ versus $\boldsymbol{R_{base}}$, revealing generalization limitations. Specifically, when switching from $\boldsymbol{R_{base}}$ to $\boldsymbol{R_{ext}}$, \texttt{Phishpedia}'s identification rate increases from 71.34\% to 85.04\% (13.7\%{\small\faLongArrowUp}), \texttt{Involution} from 81.03\% to 84.47\% (3.44\%{\small\faLongArrowUp}), and \texttt{PhishZoo} from 9.86\% to 28.73\% (18.87\%{\small\faLongArrowUp}). This indicates that logo variations significantly enhance \texttt{Phishpedia} and \texttt{Involution}'s identification capabilities. Conversely, \texttt{PhishIntention}, \texttt{VisualPhishNet}, and \texttt{EMD} maintain stable performance across reference changes, attributed to CRP-based filtering and screenshot-based methods' inherent resilience to limited reference variations.

% \observ{The discrepancy underscores a critical gap in the models' ability to reliably identify targeted brands in phishing attempts, revealing a potential weakness of the models}

%===================================
\subsection{In-depth Analysis of Detection Failures}
\label{sec:In-depth Analysis of Detection Failures}

\begin{table}[!t]
    \centering
    \caption{\textbf{Failure Statistics of Different Evasion Strategies.}}
    \label{tab:manipulation_statistics}
    %\vspace{-5px}
    \resizebox{\linewidth}{!}{
    \begin{NiceTabular}{p{.11\linewidth} p{.23\linewidth} r r r r}
    \toprule
    \multicolumn{2}{c}{\textbf{Strategy}} &	\multicolumn{1}{c}{\textbf{PhishIntention}}     &	 \multicolumn{1}{c}{\textbf{Phishpedia}} & \multicolumn{1}{c}{\textbf{PhishZoo}} & \multicolumn{1}{c}{\textbf{Involution}} \\ \midrule
    \multirow{3}{*}{\rotatebox[origin=c]{90}{Similar}} & WrongLogoArea & 68 \phantom{0}(6.8\%) & 19 \phantom{0}(1.9\%)  & 16 \phantom{0} (1.6\%)  & 173\phantom{0}(17.3\%) \\ 
    & CorrectLogoArea & 377 (37.7\%) & 245 (24.5\%) & 633\phantom{0} (63.3\%)& 67(6.7\%)\\ \cmidrule{2-6}
    & Total & 445    	(44.5\%)&   	264 	(26.4\%)&   	   649    (64.9\%)&	       240    (24.0\%) \\ \midrule %18\phantom{0}(1.80\%) 632 (63.20\%)

    \multirow{19}{*}{\rotatebox[origin=c]{90}{Visible}} & Elimination&          44 \phantom{0}(4.4\%)&    	88    	\phantom{0}(8.8\%)&   	   37     	\phantom{0}(3.7\%)&	       48     	\phantom{0}(4.80\%) \\
    & BrokenImage&          1  	    \phantom{0}(0.1\%)&    	1 	    \phantom{0}(0.1\%)&    	   3      	\phantom{0}(0.3\%)&	       1      	\phantom{0}(0.1\%) \\
    & ColorReplace&          	    52     	\phantom{0}(5.2\%)&   	58    	\phantom{0}(5.8\%)&   	   45     	\phantom{0}(4.5\%)&	       48     	\phantom{0}(4.8\%) \\
    & LogoBackground&         	26  	    \phantom{0}(2.6\%)&    	157    	(15.7\%)&    	   26     	\phantom{0}(2.6\%)&	       7      	\phantom{0}(0.7\%) \\
    & ImageBackground&         	28  	    \phantom{0}(2.8\%)&    	0    	\phantom{0}(0.0\%)&    	   4     	\phantom{0}(0.4\%)&	       71      	\phantom{0}(7.1\%) \\
    & Popup/Blurring&          	    63     	\phantom{0}(6.3\%)&    	36    	\phantom{0}(3.6\%)&    	   2      	\phantom{0}(0.2\%)&	       37     	\phantom{0}(3.7\%) \\
    % Blurred&            	0  	    \phantom{0}(0.00\%)&    	0 	    \phantom{0}(0.00\%)&    	   3      	\phantom{0}(0.60\%)&	       1      	\phantom{0}(0.20\%) \\
    & Integration&            63  	\phantom{0}(6.3\%)&    	54    	\phantom{0}(5.4\%)&    	   48     	\phantom{0}(4.8\%)&	       117     	(11.7\%) \\
    & Re-position&           	43     	\phantom{0}(4.3\%)&    	36    	\phantom{0}(3.6\%)&    	   12      	\phantom{0}(1.2\%)&	       56      	\phantom{0}(5.6\%) \\
    & Outdated&           	154     	(15.4\%)&    	194    	(19.4\%)&    	   8     	\phantom{0}(0.80\%)&	       102    	(10.2\%) \\
    & CaseConversion&           	    9  	    \phantom{0}(0.9\%)&    	32    	\phantom{}(3.2\%)&    	   15      	\phantom{0}(1.5\%)&	       43     	\phantom{0}(4.3\%) \\
    & TextAsLogo&         	5  	    \phantom{0}(0.5\%)&    	14    	\phantom{0}(1.4\%)&    	   9   	\phantom{0}(0.9\%)&	       3      	\phantom{0}(0.3\%) \\
    & ScalingOrResizing&         8   	\phantom{0}(0.8\%)&    	0 	    \phantom{0}(0.0\%)&    	   18      	\phantom{0}(1.8\%)&	       4      	\phantom{0}(0.4\%) \\
    & FontReplace&           	    4  	    \phantom{0}(0.4\%)&    	9 	    \phantom{0}(0.9\%)&    	   1      	\phantom{0}(0.1\%)&	       8      	\phantom{0}(0.80\%) \\
    & Omission&           	7  	    \phantom{0}(0.7\%)&    	23 	    \phantom{0}(2.3\%)&    	   38      	\phantom{0}(3.8\%)&	       32      	\phantom{0}(3.20\%) \\
    & Shape&          	    3  	    \phantom{0}(0.3\%)&    	16 	    \phantom{0}(1.6\%)&    	   0      	\phantom{0}(0.0\%)&	       18      	\phantom{0}(1.8\%) \\
    & ImageAddText&           36   	\phantom{0}(3.6\%)&    	13 	    \phantom{0}(1.3\%)&    	   79      	\phantom{0}(7.9\%)&	       153      	(15.3\%) \\
    & LogoAddText&           2   	\phantom{0}(0.2\%)&    	4 	    \phantom{0}(0.4\%)&    	   0      	\phantom{0}(0.0\%)&	       7      	\phantom{0}(0.7\%) \\
    & Replacement&            0   	\phantom{0}(0.0\%)&    	1 	    \phantom{0}(0.1\%)&    	   6      	\phantom{0}(0.6\%)&	       5      	\phantom{0}(0.5\%) \\
    & Blocked&       1   	\phantom{0}(0.1\%)&    	0 	    \phantom{0}(0.0\%)&    	   0      	\phantom{0}(0.0\%)&	       0      	\phantom{0}(0.0\%) \\ \cmidrule{2-6}
    & Total &       555   	(55.5\%)&    	736	    (73.6\%)&    	   351    	(35.1\%)&	       760  	(76.0\%) \\ \midrule
    \multicolumn{2}{l}{Total} &       1,000   	\phantom{0}(100\%)&    	1,000 	    \phantom{0}(100\%)&    	   1,000      	\phantom{0}(100\%)&	       1,000      	\phantom{0}(100\%) \\
    \bottomrule
    \end{NiceTabular}
    }
    %\vspace{-5px}
\end{table}
\begin{table}[t]
   \caption{\textbf{Analysis of Detection Failures for Visually Similar Phishing Logos in `CorrectLogoArea.'}}
   \label{tab:perturbation_statistics}
   \centering
   \resizebox{\linewidth}{!}{
   \begin{NiceTabular}{lrrlrr}
       \toprule
       \multicolumn{1}{c}{\textbf{Model}} & \multicolumn{1}{c}{\textbf{T.$^*$}} & \multicolumn{1}{c}{\#\textbf{Sample}} & \multicolumn{1}{c}{\textbf{Feat. Sim.$^{**}$}} & \multicolumn{1}{c}{\textbf{SSIM ($<$ 0.7)}} & \multicolumn{1}{c}{\textbf{PSNR ($>$ 4)}}\\ 
       \midrule
       PhishIntention & 0.83 & 28 & 27 (0.6--0.83)$^{***}$ &  27 & 11\\
       Phishpedia & 0.83 & 206 & 190 (0.6--0.83) & 206& 205\\
       PhishZoo & 40 & 633 & 633 (0--40) & 569 & 628\\
       Involution & 0.7 & 67 & 62 (0.6--0.7) & 66  & 60\\
       \bottomrule
       \multicolumn{6}{l}{$^*$: Threshold. $^{**}$: Feature Similarity.  $^{***}$: One sample is misclassified.}\\
   \end{NiceTabular}}
\end{table}

We find that real-world phishing attackers frequently modify four main visual components (logo, popup, login, and others) to evade phishing detection systems. These strategies and descriptions are summarized in \autoref{tab:pass_screenshot_category} of \aref{sec:pass_example_categorization}.
To quantify the prevalence of evasion strategies, we conduct a systematic manual review of 6,000 phishing samples from $D_{learn}$ that evaded detection, randomly selecting 1,000 failed samples from each model. We exclude \texttt{DynaPhish} due to its structural similarity to \texttt{PhishIntention}.
\looseness=-1

In our examination, we observe varying degrees of visual modifications, from obvious alterations detectable by humans to subtle changes that are challenging to identify.
% Some manipulated logos and screenshots display obvious alterations detectable by humans, while others feature subtle changes that prove more challenging to identify. 
To categorize these modifications, we establish two primary classification terms: `similar' and `visible.' For samples classified as `similar,' the models successfully identify the correct logo placement designated as `CorrectLogoArea,' while incorrect placements are termed `WrongLogoArea.'
The comprehensive results of our logo-based model analysis are presented in~\autoref{tab:manipulation_statistics}, with detailed performance matrices for `CorrectLogoArea' documented in~\autoref{tab:perturbation_statistics}.

\subsubsection{Logo-based Methods} 
\autoref{tab:manipulation_statistics} reveals that 44.5\%, 26.4\%, 64.9\%, and 24.0\% of failed samples appear similar logos to brand target lists for \texttt{PhishIntention}, \texttt{Phishpedia}, \texttt{PhishZoo}, and \texttt{Involution}, respectively. 
Among these, 6.8\%, 1.9\%, 1.6\%, and 17.3\% of samples fail to locate accurate logos (`WrongLogoArea') with the top-1 predicted boundary box without additional filtering.

\PP{Analysis of Similarity-based Evasion} 
We further analyze phishing samples that visually mimic legitimate target logos and where models correctly identify logo regions but fail in detection (`CorrectLogoArea' failures). Samples failing to bypass validations (CRPs and logo ratio checks) are excluded. The remaining sample sizes are as follows:
\texttt{PhishIntention} (28 samples), \texttt{Phishpedia} (206 samples), \texttt{PhishZoo} (633 samples), and \texttt{Involution} (67 samples). Most of the feature similarities between samples and reference lists fall just below each model's detection threshold---the majority of \texttt{PhishIntention} and \texttt{Phishpedia} samples at 0.6--0.83 (threshold: 0.83), all samples of \texttt{PhishZoo} at 0--40 (threshold: 40), and most \texttt{Involution} samples between 0.6--0.7 (threshold: 0.7).

\begin{figure}[!t]
    \centering
    \begin{subfigure}{0.2\textwidth}
    \includegraphics[width=\textwidth]{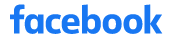} %,height=2em
    \caption{Phishing Logo}
    \label{fig:lime_phish}
    \end{subfigure}\hfill
    \begin{subfigure}{0.22\textwidth}
    \includegraphics[width=\textwidth]{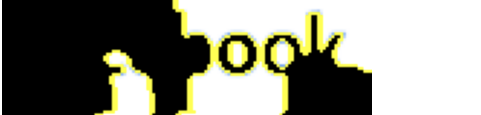} %,height=2.5em
    \caption{LIME Analyzed Logo}
    \label{fig:lime_analysis}
    \end{subfigure}\hfill
    \vspace{-5px}
    \caption{Analysis of a Facebook phishing logo using LIME. Black means negative or no contributions.}
    \label{fig:lime}
    \vspace{-10px}
\end{figure}

The visual quality metrics SSIM~\cite{Zhou2004SSIM} and PSNR~\cite{Alain2010PSNR} suggest that these samples maintain a reasonable visual similarity to legitimate logos, although the specific target logo images are unknown and calculations are based on researchers' selection. Additionally, a small subset of samples demonstrates high quality with high values of SSIM and PSNR.
%
% The visual quality metrics SSIM~\cite{Zhou2004SSIM} and PSNR~\cite{Alain2010PSNR} suggest these samples maintain high visual similarity to legitimate logos. High PSNR values across models (\texttt{PhishZoo}: 564 samples at 8--16, \texttt{Phishpedia}: 158 samples at 4--6) indicate preserved visual quality. 
LIME~\cite{lime} analysis of a Facebook phishing example (~\autoref{fig:lime}) provides additional insight: while the logo appears authentic, modifications appear in areas that may affect model detection but remain less noticeable to humans.
% 
% These patterns, further illustrated in~\autoref{fig:failure_example}, 
These findings could indicate potential adversarial manipulation in real-world phishing attacks, suggesting attackers might be developing methods to maintain visual similarity while avoiding detection thresholds.
\looseness=-1

\PP{DynaPhish, PhishIntention, and Phishpedia} 
The systems \texttt{PhishIntention} and \texttt{DynaPhish} share a core structure of logo detection, brand verification, and domain checking, with some distinctions: \texttt{PhishIntention} adds CRP checking and OCR-aided features, while \texttt{DynaPhish} employs Google Search APIs for brand-domain verification. Several vulnerabilities exist in these approaches: Faster-RCNN's imprecise logo detection (as seen in \autoref{fig:c}), the assumption that phishing sites require CRPs (missing alternatives like QR codes), and challenges with borderline logo similarity cases (\autoref{fig:d}). Attackers can exploit these weaknesses by manipulating URLs or adjusting brand similarities. For non-targeted brand logos, attackers maintain semantic meaning while controlling similarity through strategies like 'Elimination' (\eg, \autoref{fig:e}), background modifications, or text additions. The OCR integration in \texttt{PhishIntention} improves textual logo detection, as demonstrated in \autoref{fig:f}, where it successfully identifies Facebook (similarity: 0.92) while \texttt{Phishpedia} fails (similarity: 0.78, below its 0.83 threshold). Additional examples appear in \autoref{fig:failure_example}.

\PP{Involution}
The logo area detection and Involution model are used as the pipeline. \autoref{tab:manipulation_statistics} reveals that only 24\% of samples have matching logos, while 76\% bypass detection through basic modifications. Specifically, logo removal causes 4.80\% of failures, text additions to screenshots impair Faster-RCNN's logo region detection in 15.30\% of cases, and alternative logo integration (\eg, \autoref{fig:i}) accounts for 11.7\% of failures.

\PP{PhishZoo}
The system shows strong phishing detection but has limitations in identification due to SIFT's poor performance in matching screenshots and logos. The keyword selection process from parsed URLs and HTML is problematic -- for example, in \autoref{fig:g}, it selects generic terms like ``Page'' and ``Password,'' while for legitimate AT\&T content, it chooses irrelevant words like ``arrowmenu'' and ``verse.'' Neither set captures phishing-specific markers, reducing accuracy. Additionally, 35.10\% of failures occur when samples differ from target appearances, with text additions significantly impacting results, as shown in \autoref{fig:h}.

\subsubsection{Screenshot-based Methods} 
Identifying the exact cause of failure is challenging due to the multiple visual components in screenshots. Therefore, we examine distance or similarity metrics. We find that 93.40\% of failed samples for \texttt{EMD} have distances ranging from 0.90 to 0.94, close to its detection threshold (0.94), indicating that these samples are particularly challenging to distinguish without additional context. Additionally, 64.40\% of failed samples fall within the 0.93 to 0.94 range, while 6.60\% display zero EMD values from the target list. \autoref{fig:a} illustrates an example where the EMD value is zero, and the target reference list only contains outdated screenshots. 
For \texttt{VisualPhishNet}, distances vary from 1.00 to 1.99, with 83.30\% of samples having distances between 1.00 and 1.50 and only 16.70\% ranging from 1.5 to 2.0.
\autoref{fig:b} shows an example very similar to a screenshot in the reference list but slightly above the threshold (1.0). These observations suggest that setting fixed thresholds can be risky because attackers design adversarial images to closely mimic benign ones while slightly exceeding the threshold of detectors.
\looseness=-1

\observ{Analysis reveals three key weaknesses in current phishing detection systems: (1) pipeline exploitation through logo manipulation and CRP circumvention, (2) visually plausible modifications that remain convincing to humans while falling below detection thresholds, and (3) straightforward evasion techniques such as text overlay and logo removal. These findings indicate an overreliance on static feature matching and predetermined thresholds, underscoring the necessity for detection methods that incorporate dynamic, contextual awareness}

% \vspace{-5px}

%======================

\subsection{Ablation Study} 
\label{sec:Ablation_Study}

\begin{table}[!t]
    \renewcommand{\arraystretch}{1}
    \setlength{\tabcolsep}{2pt}
    \centering
    \normalsize
    \caption{\textbf{Ablation Study Result.}}
    \label{tab:ablation_results}
    \vspace{-5px}
    \resizebox{\columnwidth}{!}{
        \begin{tabular}{lcccrr}
        \toprule
           \multirow{2}{*}{\makecell[c]{\textbf{Model}}} &  \multicolumn{3}{c}{\textbf{Component}}  & \multicolumn{2}{c}{\textbf{Result}} \\
            \cmidrule(lr){2-4} \cmidrule(lr){5-6} %\cmidrule(lr){6-6}
            &Logo & URL & HTML & \multicolumn{1}{c}{\textbf{Detection}} & \multicolumn{1}{c}{\textbf{Identification}}\\ \midrule
            
            \multirow{4}{*}{\textbf{PhishIntention}} &  \textbf{C$^1$}&  \textbf{C}&  \textbf{C} & 3/110 (2.73\%)&3/3 (100.00\%)\\
            &  \textbf{C}&  \textbf{C}&  \textbf{M$^2$}  & 2/110 (1.82\%)& 2/2 (100.00\%)\\
            &  \textbf{M}&  \textbf{C}&  \textbf{C} & 358/7,113 (5.03\%) & 165/358 (46.09\%) \\
            &  \textbf{C}&  \textbf{M}&  \textbf{C} & 296/1,321 (22.41\%) & 296/296 (100.00\%)\\ \midrule
            \multirow{4}{*}{\textbf{Phishpedia}} &  \textbf{C}&  \textbf{C}&  \textbf{C} & 8/110 (7.27\%)& 7/8 (87.50\%)\\
            &  \textbf{C}&  \textbf{C}&  \textbf{M} & ---- & ----\\
            &  \textbf{M}&  \textbf{C}&  \textbf{C} & 991/7,113 (13.93\%) & 323/991 (32.59\%)\\
            &  \textbf{C}&  \textbf{M}&  \textbf{C} & 906/1,321 (68.58\%) & 894/906 (98.69\%) \\ \midrule
            \multirow{4}{*}{\textbf{DynaPhish}} &  \textbf{C}&  \textbf{C}&  \textbf{C} & 3/110 (2.73\%) & 3/3 (100.00\%) \\
            &  \textbf{C}&  \textbf{C}&  \textbf{M} & 2/110 (1.82\%) & 2/2(100.00\%)\\
            &  \textbf{M}&  \textbf{C}&  \textbf{C} & 1,372/7,113 (19.29\%) & 1,218/1,372 (88.78\%)\\
            &  \textbf{C}&  \textbf{M}&  \textbf{C} & 77/1,321 (5.83\%) & 77/77 (100.00\%)\\ \midrule
            \multirow{4}{*}{\textbf{PhishZoo}} &  \textbf{C}&  \textbf{C}&  \textbf{C} & 81/110 (73.64\%) & 4/81 (4.94\%)\\
            &  \textbf{C}&  \textbf{C}&  \textbf{M} & 24/110 (21.82\%) & 3/24 (12.50\%)\\
            &  \textbf{M}&  \textbf{C}&  \textbf{C} & 6,916/7,113 (97.23\%) & 1,472/6,916 (21.28\%)\\
            &  \textbf{C}&  \textbf{M}&  \textbf{C} & 973/1,321 (73.66\%) & 48/973 (4.93\%)\\ \bottomrule
            
            \multicolumn{6}{l}{\makecell[l]{
                $^1$\textbf{$C$} = Controlled  component;$\>\>$ $^2$\textbf{$M$} = Modified Components: manipulated logos \\ typo-squatting URLs,  and empty HTMLs.}}
         \end{tabular}}
        \vspace{-10px}
\end{table}

% \begin{table}[]
% \begin{tabular}{clllll}
% \multirow{2}{*}{Model}    & \multicolumn{3}{l}{Choose} & \multicolumn{2}{l}{RRRR} \\
%                           & A       & B       & C      &             &            \\
% \multicolumn{1}{l}{CCCCC} &         &         &        &             &           
% \end{tabular}
% \end{table}
% We further examine how the models (\texttt{PhishIntention}, \texttt{Phishpedia}, \texttt{DynaPhish}, and \texttt{PhishZoo}) leverage multiple data components (specifically URLs, HTML structure, and visual screenshots) to make accurate classifications. To rigorously evaluate the importance of visual features, we design controlled experiments that systematically vary the URL and HTML components while isolating the visual elements.

% To rigorously assess the impact of multiple data components—URLs, HTML structure, and visual screenshots—we design controlled experiments involving models (\texttt{PhishIntention}, \texttt{Phishpedia}, \texttt{DynaPhish}, and \texttt{PhishZoo}).

\PP{Study Plan}
We conduct a detailed ablation study evaluating the performance of four models (\texttt{PhishIntention}, \texttt{Phishpedia}, \texttt{DynaPhish}, and \texttt{PhishZoo}) across distinct scenarios. The first scenario uses completely legitimate inputs, incorporating authentic logos, URLs, and HTML structure. This establishes our baseline for normal website behavior. In the second scenario, we modify the input by maintaining legitimate logos and HTML while introducing typosquatted URLs, allowing us to assess the impact of URL manipulation. The third scenario uses legitimate logos and URLs but combines them with an empty HTML structure, enabling us to isolate the role of HTML content in detection accuracy.

\PP{Results}
\autoref{tab:ablation_results} presents the performance of phishing detection and identification in our ablation study for \texttt{PhishIntention}, \texttt{Phishpedia}, \texttt{DynaPhish}, and \texttt{PhishZoo}. 
In \autoref{tab:ablation_results}, controlled original legitimate components are denoted as \textbf{C} and modified components as \textbf{M}.

Our analysis comparing configurations with and without HTML components (`CCC' and `CCM') reveals varying impacts on false positive rates. \texttt{PhishIntention} and \texttt{DynaPhish} showed a minor improvement, reducing false positives from 3 to 2 cases due to their handling of inadequately maintained domains. More notably, \texttt{PhishZoo} demonstrated substantial improvement, with false positives decreasing from 81 to 24 when HTML was removed, suggesting that its HTML analysis through keyword matching may impair detection accuracy and require methodology refinement.

The analysis reveals significant variations in phishing detection performance when comparing original versus modified logos (`CCC' and `MCC'). \texttt{PhishIntention}, \texttt{Phishpedia}, and \texttt{DynaPhish} showed relatively low detection rates of 5.03\%, 13.93\%, and 19.29\% respectively with modified logos, indicating substantial vulnerability to logo-based evasion tactics. While \texttt{PhishZoo} reported a 97.23\% detection rate, this figure appears inflated due to poor identification accuracy and high false positives. \texttt{DynaPhish}'s superior performance in detecting modified logos (19.29\% vs \texttt{PhishIntention}'s 5.03\%) can be attributed to its real-time web search capability, demonstrating the value of incorporating online search features in phishing detection systems.

Under modified URL configurations (`CCC' and `CMC'), \texttt{Phishpedia} achieves a 68.58\% detection rate, demonstrating both resilience and limitations in its dual URL-logo analysis approach. This performance indicates the need to enhance domain verification through expanded databases or alternative domain-brand authentication methods.

\observ{Logos, URLs, and HTML are critical components that significantly influence the results. The current simple processing of HTML for \texttt{PhishZoo} has detrimental impacts on its performance}

%==============================
\section{Evaluation of Model Robustness against Manipulated Visual Components} \label{sec:robustness_eval}

\begin{table*}[!t]
    \renewcommand{\arraystretch}{1.05}
    \setlength{\tabcolsep}{1.4pt}
    \centering
    \normalsize
    \caption{\textbf{Evaluation Results of Phishing Detection with Manipulated Visual Components.} Models, trained on $\boldsymbol{R_{ext}}$, are used for evaluation. 
    The row means manipulating type; the original is the brand's recent benign webpage, visible manipulation, and perturbation-based manipulation, which refers to the dataset crafted for robustness testing. The column for different models means URL type, benign refers to the original benign URL, while squatted means the created URLs. Default benign URLs are used. Note that the row of `Original' with benign URLs means \textit{misclassified}.}
    \label{tab:result_making_detection}
    \vspace{-5px}
    \resizebox{\linewidth}{!}{
        \begin{tabular}{lp{0.11\linewidth}rrrrrrrrrr}
        \toprule
          \multirow{2}{*}{\textbf{}} & \multirow{2.5}{*}{\makecell[c]{\textbf{Manipulation}\\\textbf{Name}}} &  \multicolumn{2}{c}{\textbf{PhishIntention}}  & \multicolumn{2}{c}{\textbf{Phishpedia}} & \multicolumn{1}{c}{\multirow{2}{*}{\textbf{DynaPhish}}} & 
          \multicolumn{1}{c}{\multirow{2}{*}{\textbf{Involution}}} & \multicolumn{1}{c}{\multirow{2}{*}{\textbf{PhishZoo}}} & \multirow{2}{*}{\textbf{VisualPhishNet}} & \multicolumn{1}{c}{\multirow{2}{*}{\textbf{EMD}}} \\ \cmidrule(lr){3-4} \cmidrule(lr){5-6}
          &  &  \multicolumn{1}{c}{Benign} & \multicolumn{1}{c}{Squatted} & \multicolumn{1}{c}{Benign} & \multicolumn{1}{c}{Squatted} &  &  &  &  & \\ \midrule
&Original & 3/110\phantom{0 }(2.73\%) & 316/1,321 (23.92\%) & 8/110\phantom{0 }(7.27\%) & 894/1,321 (67.68\%) & 3/110 (2.73\%) & 8/110 (7.27\%)& 81/110 (73.64\%) & 30/110 (27.27\%) & 55/110 (50.00\%)\\ 
% &Original & 3/110\phantom{0 }(2.73\%) & 316/1,321 (23.92\%) & 7/110\phantom{0 }(6.36\%) & 894/1,321 (67.68\%) &  \multicolumn{1}{c}{---} & \multicolumn{1}{c}{---} & \multicolumn{1}{c}{---} & \multicolumn{1}{c}{---}\\ 
          \midrule
          \multirow{14}{*}{\rotatebox[origin=c]{90}{Visible Manipulation}}
&Elimination  & 0/110\phantom{00~}(0.0\%)     & 10/1,321\phantom{0 }(0.76\%) & 10/110\phantom{0 }(9.09\%) & 151/1,321 (11.43\%)& 1/110 (0.91\%)& 3/110\phantom{0 }(2.73\%)& 103/110 (93.64\%) & 28/110 (25.45\%) &54/110 (49.09\%) \\ 
&ColorReplace     & 4/580\phantom{0 }(0.69\%)  & 858/6,965 (12.32\%)& 44/580\phantom{0 }(7.59\%) & 1,248/6,965 (17.92\%)& 56/580 (9.66\%)&327/580 (56.38\%)& 562/580 (96.90\%) & 125/580 (21.55\%) & 275/580 (47.41\%)\\
&Resizing         & 29/877\phantom{0 }(3.30\%) & 2,230/10,532 (21.17\%)& 65/877\phantom{0 }(7.41\%) & 6,383/10,532 (60.61\%) &200/877 (22.81\%)&696/877 (79.36\%)& 856/877 (97.61\%)  & 242/877 (27.59\%) & 421/877 (48.00\%)\\
&Rotation     & 36/1,320\phantom{0 }(2.73\%) & 3,612/15,852 (22.79\%)&96/1,320\phantom{0 }(7.27\%) & 10,442/15,852 (65.87\%) & 326/1,320 (24.70)&1,085/1,320 (82.20\%)& 1,283/1,320 (97.20\%) & 387/1,320 (29.32\%) & 671/1,320 (50.83\%)\\
&Integration & 22/369\phantom{0 }(5.96\%) & 817/4,431 (18.44\%)  & 58/369 (15.72\%) & 2,624/4,431 (59.22\%) &77/369 (20.88) &217/369 (58.81\%)& 362/369 (98.10\%) & 95/369 (25.75\%) & 188/369 (50.95\%)\\
&Re-position      & 24/879\phantom{0 }(2.73\%)& 2,035/10,556 (19.28\%)& 66/879\phantom{0 }(7.51\%) & 6,278/10,556 (59.47\%)& 190/879 (21.62\%)&587/879 (66.78\%)& 870/879 (98.98\%) & 216/879 (24.57\%) & 432/879 (49.15\%)\\
&Flipping      & 4/220\phantom{0  }(1.82\%) & 459/2,642 (17.37\%)&16/220\phantom{0 }(7.27\%) & 1692/2,642 (64.04\%)&36/220 (16.36\%)&28/220 (12.73\%)&  216/220 (98.18\%)& 62/220 (28.18\%)& 107/220 (48.64\%) \\
&Replacement  & 177/1,006 (17.59\%)&2,202/12,081 (18.23\%)& 444/1,006 (44.14\%)& 5,545/12,081 (45.90\%) &145/1006 (14.41\%)&502/1,006 (49.90\%)& 967/1,006 (96.12\%)& 235/1,006 (23.36\%)& 468/1,006 (46.52\%)\\
&Blurring      & 1/110\phantom{0 }(0.91\%) & 122/1,321\phantom{0 }(9.24\%)& 3/110\phantom{0 }(2.73\%)& 426/1,321 (32.20\%) &7/110 (6.36\%)&32/110 (29.09\%)& 102/110 (92.73\%)  & 29/110 (26.36\%)&51/110 (46.36\%)\\
% &9. Blurring      & 163/220 (74.09\%) & 122/1,321\phantom{0 }(9.24\%)& 3/110\phantom{0 }(2.73\%)& 426/1,321 (32.20\%) &32/110 (29.09\%)& 102/110 (92.73\%)  & 29/110 (26.36\%)&51/110 (46.36\%)\\
&Scaling       & 20/550\phantom{0 }(3.64\%) & 1,530/6,605 (23.16\%)&41/550\phantom{0 }(7.45\%) & 4,536/6,605 (68.68\%) &144/550 (26.18\%)&461/550 (83.82\%)& 538/550 (97.82\%) & 147/550 (26.73\%) &274/550 (49.82\%)\\
&Omission  & 2/96\phantom{0 }(2.08\%)& 114/1,152\phantom{0 }(9.90\%)& 11/96 (11.46\%) & 451/1,152 (39.15\%) &5/96 (5.21\%)&52/96 (54.17\%)& 93/96 (96.88\%) & 34/96 (35.42\%)& 42/96 (43.75\%)\\
&FontReplace    & 4/186\phantom{0 }(2.15\%) & 294/2,232 (13.17\%)& 23/186 (12.37\%) & 774/2,232 (34.68\%)&26/186 (13.98\%)&115/186 (61.83\%)& 179/186 (96.24\%) & 56/186 (30.11\%)& 90/186 (48.39\%)\\
&CaseConversion   & 6/225\phantom{0 }(2.67\%) & 299/2,700 (11.07\%) & 36/225 (16.00\%)  & 960/2,700 (35.56\%)&22/225 (9.78\%)&76/225 (33.78\%)&  214/225 (95.11\%) & 73/225 (32.44\%)& 113/225 (50.22\%)\\ 
\cmidrule{2-11}
&Total            & 329/6,528\phantom{0 }(5.04\%) & 14,582/78,390 (18.60\%)& 913/6,528 (13.99\%) &  41,510/78,390 (52.95\%) & 1,235/6,528 (18.92\%)&4,181/6,528 (64.05\%) & 6,345/6,528 (97.20\%) & 1,729/6,528 (26.49\%) & 3,186/6,528 (48.81\%)\\ 
          \cmidrule{2-11}
          & SRNet & 11/41 (26.83\%)&114/492 (23.17\%)& 34/41 (82.93\%) &347/492 (70.53\%) &8/41 (19.51\%)&25/41 (60.98\%) & 39/41 (95.12\%) & 11/41 (26.83\%) & 20/41 (48.78\%)\\
          \midrule
          \multirow{6}{*}{\rotatebox[origin=c]{90}{Perturbation-based}}
          & \cite{lee2023attacking}-ViT & 3/110\phantom{0 }(2.78\%) &266/1,321 (20.14\%)& 8/110\phantom{0 }(7.27\%) & 696/1,321 (52.69\%) &26/110 (23.64\%) &82/110 (74.55\%) & 106/110 (96.36\%) & 34/110 (30.91\%) & 52/110 (47.27\%)\\ [1pt]
          & \cite{lee2023attacking}-Swin & 3/110\phantom{0 }(2.73\%) &296/1,321 (22.41\%)& 10/110\phantom{0 }(9.09\%) & 820/1,321 (62.07\%)&28/110 (25.45\%) &86/110 (78.18\%)& 108/110 (98.18\%)& 34/110 (30.91\%) & 55/110 (50.00\%)\\ [1pt]
          
          & FGSM & 4/108\phantom{0 }(3.70\%) & 279/1,297 (21.51\%) &8/108\phantom{0 }(7.41\%) & 776/1,297 (59.83\%) & 25/110 (22.73\%)&80/108 (74.07\%) & 106/108 (98.15\%) & 30/108 (27.78\%) & 52/108 (48.15\%)\\ [1pt]
          & PGD & 4/108\phantom{0 }(3.70\%)& 259/1,297 (19.97\%) & 8/108\phantom{0 }(7.41\%) & 756/1,297 (58.29\%) & 25/108 (23.15\%)&79/108 (73.15\%) & 106/108 (98.15\%) & 30/108 (27.78\%) &  51/108 (47.22\%)\\ [1pt]
          & CW &  4/108\phantom{0 }(3.70\%)& 269/1,297 (20.74\%) & 10/108\phantom{0 }(9.26\%) & 790/1,297 (60.90\%) & 25/108 (23.15\%)&80/108 (74.07\%) & 106/108 (98.15\%) & 30/108 (27.78\%) & 52/108 (48.15\%)\\ [1pt]
          \cmidrule{2-11}
          & Total & 18/544\phantom{0 }(3.31\%) &1,369/6,533 (20.96\%) & 44/544\phantom{0 }(8.09\%) & 3,838/6,533 (58.75\%) & 129/544 (23.71\%)&407/544 (74.82\%) & 532/544 (97.79\%) & 158/544 (29.04\%) & 262/544 (48.16\%)\\ [1pt]
          \bottomrule
         \end{tabular}}
        \vspace{-5px}
\end{table*}

To further investigate why visual similarity-based anti-phishing models fail, we categorize possible reasons into (1) visible manipulations, where simple modifications have the possibility to be detectable by humans; and (2) perturbed manipulations, where phishing logos closely resemble target brands. Representative manipulations are selected from~\autoref{tab:manipulation_statistics} and ~\autoref{tab:pass_screenshot_category}.  
We employ several white-box and black-box attack methods on the benign screenshots to imitate the perturbed manipulations. Details can be found in~\aref{sec:perturbated_example}.
The robustness results for seven models on crafted datasets, covering phishing detection and phishing brand identification rates, are detailed in~\autoref{tab:result_making_detection} and~\autoref{tab:result_making_identification}.
\looseness=-1

\PP{Evaluation Plan}
The crafted samples are equipped with original benign URLs and HTML for evaluation by default. To assess the impact of URLs with crafted visual images, we use squatted domains (\eg, faceb{\color{red}00}k .com) to replace the original benign URLs for \texttt{PhishIntention} and \texttt{Phishpedia}, as they employ second-level domains to verify legitimacy.
The performance represents the upper bound for methods that are not equipped with domain checks.

\PP{Settings.} 
We use domains from \texttt{PhishIntention} and the $\boldsymbol{R_{ext}}$ as reference lists. 
% The number of benign samples is denoted as $N_n$, with $N_{tn}$ representing those reported as benign, and the number of correctly identifying the target brand for benign as $I_{tn}$. The detection accuracy is $I_{tn}/N_n$, the identification rate is $I_{tn}/N_{tn}$, and the identification rate out of the total is $I_{tn}/N_n$.
Metrics are the same as in~\autoref{sec:results}.

\subsection{Result: Robustness Evaluation} 
\label{subsubsec:craft_detection_results}
\PP{Legitimate Samples (False Positive)}
Legitimate samples (legitimate screenshots, URLs, and HTML) are expected to be correctly identified as benign. Incorrectly labeling benign samples as potential phishing attempts is defined as a false positive error. 
As shown in~\autoref{tab:result_making_detection}, \texttt{PhishIntention} detects 3 benign samples (with legitimate domains) as phishing, 8 for \texttt{Phishpedia}, and 3 for \texttt{DynaPhish}.
The misclassification of \texttt{PhishIntention} and \texttt{Phishpedia} stems from the reliance on brand-domain verification. Some legitimate domains, such as `santanderbank,' are not included in the list, although `santander' and `santanderresearch' exist in the list.
This oversight highlights the limitations of relying on incomplete reference lists for verification purposes. The error for \texttt{DynaPhish} comes from the ``forbidden words'' maintained by the authors.
\looseness=-1

\observ{Models that rely on incomplete reference lists and static word matching to verify the legitimacy of logos and domains are prone to false positive errors, incorrectly flagging legitimate websites as phishing threats}

\begin{table*}[!t]
    \renewcommand{\arraystretch}{1.05}
    \setlength{\tabcolsep}{1.4pt}
    \centering
    \normalsize
    \caption{\textbf{Evaluation Results of Phishing Brand Identification with Manipulated Visual Components.} Models, trained on $\boldsymbol{R_{ext}}$, are used for evaluation. The percentage is the correctly identified brands out of the predicted phishing number by default.}
    \label{tab:result_making_identification}
    \vspace{-5px}
    \resizebox{\textwidth}{!}{
    \begin{tabular}{lp{0.11\linewidth}rrrrrrrrrr}
        \toprule
          \multirow{2}{*}{\textbf{ }} & \multirow{2.5}{*}{{\makecell[c]{\textbf{Manipulation}\\\textbf{Name}}}} &  \multicolumn{2}{c}{\textbf{PhishIntention}}  & \multicolumn{2}{c}{\textbf{Phishpedia}}  & \multicolumn{1}{c}{\multirow{2}{*}{\textbf{DynaPhish}}} & \multicolumn{1}{c}{\multirow{2}{*}{\textbf{Involution}}} & \multicolumn{1}{c}{\multirow{2}{*}{\textbf{PhishZoo}}} & \multicolumn{1}{c}{\multirow{2}{*}{\textbf{VisualPhishNet}}} & \multicolumn{1}{c}{\multirow{2}{*}{\textbf{EMD}}} \\ 
          \cmidrule(lr){3-4} \cmidrule(lr){5-6}
          &  &  \multicolumn{1}{c}{Benign} & \multicolumn{1}{c}{Squatted} & \multicolumn{1}{c}{Benign} & \multicolumn{1}{c}{Squatted} &  &  &  &  & \\ 
          \midrule
          &Original & 3/3 (100.0\%) & 316/316 (100.0\%) & 7/8 (87.50\%) & 882/894 (98.66\%) & %\multicolumn{1}{c}{---} & \multicolumn{1}{c}{---} & \multicolumn{1}{c}{---} & \multicolumn{1}{c}{---}\\ 
          3/3 (100.0\%) &7/8 (87.50\%) & 4/81 (4.94\%) & 16/30 (53.33\%) & 11/55 (20.0\%)\\
          \midrule
          \multirow{14}{*}{\rotatebox[origin=c]{90}{Visible Manipulation}}            
&Elimination & 0/0\phantom{00 }(0.0\%)         &10/10 (100.0\%) & 0/10\phantom{00 }(0.0\%) & 30/151 (19.87\%)&1/1 (100.0\%) &3/3 (100.0\%)& 7/103\phantom{0 }(6.80\%)& 6/28 (25.43\%)&9/54 (16.67\%)  \\ 
&ColorReplace& 4/4 (100.0\%)       &858/858 (100.0\%)& 6/44 (13.64\%) &792/1,248 (63.46\%)& 56/56 (100.0\%) &322/327 (98.47\%) & 69/562 (12.28\%) & 39/125 (31.20\%) & 49/275 (17.82\%)\\
&Resizing       & 29/29 (100.0\%)     &2,230/2,230 (100.0\%)& 42/65 (64.62\%) & 6,112/6,383 (95.75\%)&200/200 (100.0\%) &688/696 (98.85\%)& 160/856 (18.69\%)  & 140/242 (57.85\%) & 80/421 (19.00\%)\\
&Rotation      & 36/36 (100.0\%)     &3,612/3,612 (100.0\%)& 81/96 (84.38\%) &10,262/10,442 (98.28\%)&326/326 (100\%)&1,072/1,085 (98.80\%)& 389/1,283 (30.32\%) & 202/387 (52.20\%) & 121/671 (18.03\%) \\
&Integration & 9/22 (40.91\%)    &661/817 (80.91\%)& 23/58 (39.66\%)& 2,206/2,624 (84.07\%)&61/77 (79.22\%)&188/217 (86.64\%)& 119/362 (32.87\%) & 38/95 (40.00\%) & 37/188 (19.68\%)\\
&Location & 24/24 (100.0\%)     &2,035/2,035 (100.0\%)& 47/66 (71.21\%)& 6,051/6,278 (96.38\%)&190/190 (100.0\%)&578/587 (98.47\%)& 288/870 (33.10\%) & 103/216 (47.69\%) & 79/432 (18.29\%)\\
&Flipping & 4/4 (100.0\%)       &459/459 (100.0\%)&13/16 (81.25\%)& 1,656/1,692 (97.87\%)&36/36 (100.0\%)&28/28 (100.0\%)&  28/216 (12.96\%)& 34/62 (54.84\%)& 19/107 (17.76\%) \\
&Replacement& 0/177\phantom{00 }(0.0\%)       &70/2,202\phantom{0 }(3.18\%)& 0/444\phantom{00 }(0.0\%) & 210/5,545\phantom{0 }(3.79\%)&7/145 (4.83\%)&27/502\phantom{0 }(5.38\%)& 63/967\phantom{0 }(6.51\%)& 63/235 (26.81\%)& 67/468 (14.32\%)\\
&Blurring & 1/1 (100.0\%)       &122/122 (100.0\%)& 1/3 (33.33\%)& 402/426 (94.37\%)&7/7 (100.0\%)&23/32 (71.88\%)& 5/102\phantom{0 }(4.90\%)  & 17/29 (58.62\%)&11/51 (21.57\%) \\
&Scaling& 20/20 (100.0\%)     &1,530/1,530 (100.0\%)&35/41 (85.37\%) & 4,466/4,536 (98.46\%)&144/144 (100.0\%)&455/461 (98.70\%)& 155/538 (28.81\%) & 89/147 (60.54\%) &48/274 (17.52\%) \\
&Omission& 2/2 (100.0\%)       &114/114 (100.0\%)& 3/11 (27.27\%)& 355/451 (78.71\%)& 5/5 (100.0\%)&47/52 (90.38\%)& 20/93 (21.51\%) & 8/34 (23.53\%)& 5/42 (11.90\%)\\
&FontReplace& 4/4 (100.0\%)       &294/294 (100.0\%)& 5/23 (21.74\%) & 558/774 (72.09\%)& 26/26 (100.0\%)&111/115 (96.52\%)& 24/179 (13.41\%) & 25/56 (44.64\%)& 17/90  (18.89\%)\\
&CaseConversion& 3/6\phantom{0 }(50.0\%)        &263/299 (87.96\%) & 5/36 (13.89\%)& 588/960 (61.25\%)& 22/22 (100.0\%)&74/76 (97.37\%)& 25/214 (11.68\%) & 36/73 (49.32\%)& 20/113 (17.70\%)\\
\cmidrule{2-11}
&Total  & 136/329 (41.34\%) &12,258/14,582 (84.06\%) & 261/913 (28.59\%) & 33,688/41,510 (81.16\%)&1,081/1,235 (87.53\%)&3,616/4,181 (86.49\%)& 1,352/6,345 (21.31\%) & 800/1,729 (46.27\%) & 562/3,186 (17.64\%) \\ 
\cmidrule{2-11}
& SRNet & 11/11 (100.0\%)     &114/114 (100.0\%)& 34/34 (100.0\%) &347/347 (100.0\%)&8/8 (100.0\%)&25/25 (100.0\%)& 6/39 (15.38\%) & 5/11 (45.45\%) & 3/20 (15.00\%)  \\
\midrule
\multirow{6}{*}{\rotatebox[origin=c]{90}{Perturbation-based}} 
& \cite{lee2023attacking}-ViT   & 3/3 (100.0\%)       &266/266 (100.0\%)& 4/8 (50.00\%)& 648/696 (93.10\%)&26/26 (100.0\%)&81/82 (98.78\%)& 27/106 (25.47\%) & 18/34 (52.94\%) & 10/52 (19.23\%) \\ [1pt]
& \cite{lee2023attacking}-Swin  & 3/3 (100.0\%)       &296/296 (100.0\%)& 6/10 (60.00\%) & 772/820 (94.15\%)&28/28 (100.0\%)&85/86 (98.84\%)& 35/108 (32.41\%) & 17/34 (50.00\%) & 12/55 (21.82\%) \\ [1pt]
          & FSGM &4/4 (100.0\%)& 279/279 (100.0\%)& 6/8 (75.00\%) & 752/776 (96.91\%) & 25/25 (100\%) &79/80 (98.75\%) & 19/106 (17.92\%) & 15/30 (50.00\%) & 9/52 (17.31\%)\\ [1pt]
          & PGD &4/4 (100.0\%)& 259/259 (100.0\%)& 6/8 (75.00\%) & 732/756 (96.83\%) & 25/25 (100\%)&78/79 (98.73\%) & 14/106 (13.21\%)& 14/30 (46.67\%) &  7/51 (13.73\%)\\ [1pt]
          & CW &4/4 (100.0\%)& 269/269 (100.0\%)& 6/10 (60.00\%) & 742/790 (93.92\%) & 25/25 (100\%)&79/80 (98.75\%) & 19/106 (17.92\%)& 15/30 (50.00\%) & 9/52 (17.31\%)\\ [1pt]
          \cmidrule{2-11}
          & Total & 18/18 (100.0\%) & 1,369/1,369 (100.0\%) & 28/44 (63.64\%) & 3,646/3,838 (95.00\%) & 129/129 (100.0\%)&402/407 (98.77\%) & 114/532 (21.43\%) & 79/158 (50.00\%) & 47/262 (17.94\%)\\ [1pt]
          \bottomrule
    \end{tabular}}
    \vspace{-5px}

\end{table*}

\PP{Visible and Perturbation-based Manipulation Methods} 
From~\autoref{tab:result_making_detection} and~\autoref{tab:result_making_identification}, we observe that visible and perturbation-based strategies impact the identification result, but they do not affect the detection result significantly for \texttt{PhishZoo}. Specifically, it achieves a 97.79\% detection rate on the perturbation-based adversarial manipulation dataset and 96.16\% on the visible manipulation dataset. Logo elimination, blurring, replacing font manually or by \texttt{SRNet}, and converting cases are critical factors in the model's phishing detection capability, whereas replacing logos markedly influences identification results. 
\texttt{PhishZoo} is less sensitive to the combined logos but sensitive to the white-box attack in phishing identification, which means it is not robust on the perturbation-based manipulations. For instance, Vit and Swin-based methods achieve 25.47\% and 32.41\% while FSGM, PGD, and CW only achieve 17.92\%, 13.21\%, and 17.92\%, respectively. 
For other logo-based approaches, manipulations such as logo deletion, flipping, blurring, and case conversion substantially affect detection results. Meanwhile, changing colors, combining logos, or replaced with other logos play important roles in identification. Although logo text font greatly affects \texttt{Phishpedia}, it does not significantly impact \texttt{Involution}. \texttt{DynaPhish} is effective and robust in recognizing brands but may make mistakes when there are two logos.

Although the influence of perturbation-based attacks is not as great as the visible manipulation, they reveal weaknesses in the models: \texttt{PhishIntenion} and \texttt{Phishpedia} are sensitive to the \texttt{ViT-based} black-box attack and the \texttt{PGD} white-box attack. \texttt{Involution} is not robust on white-box attacks, and \texttt{PhishZoo} is susceptible to both attacks. For screenshot-based methods, detection performance remains stable across various manipulations, with perturbation-based manipulation strategies even improving detection rates in \texttt{VisualPhishNet}. However, these methods struggle with accurately identifying the target brand. Additionally, decreased performance observed when logos are deleted, replaced, or divided in identification results underscores the crucial role of logos in brand recognition within screenshot-based methods. 
We mention that the attacked logos are obtained based on one model and transferred to test other models. The results indicate the transferability of the attacks.

\observ{Simple visible and perturbation-based manipulations significantly disrupt logo-based methods. Both of them are transferable.
Screenshot-based methods maintain stable detection but struggle with identifying brands when logos are altered}

\PP{Benign Vs. Squatted Domains}
\texttt{PhishIntention} shows varied performance shifts when tested with squatted versus benign URLs: +13.56\% for manual visible manipulations, -3.66\% for SRNet, and +17.65\% for perturbation-based manipulations. \texttt{Phishpedia}'s detection rates change from 13.99\% to 52.95\% for manual visible manipulations, 82.93\% to 70.5\% for SRNet, and 8.09\% to 58.75\% for perturbation-based manipulations.
\looseness=-1

These results highlight the critical role of domain validation in both models' detection mechanisms. By comparing detected brand domains with parsed URL domains, the models become vulnerable to URL manipulation attacks. Attackers can potentially bypass detection by using squatted domains that match benign domains (e.g., `www.capitalone.aaa' targeting `www.capitalone.com'). URL structure parsing also presents vulnerabilities, as demonstrated by \texttt{tldextract} (\url{https://github.com/john-kurkowski/tldextract}) parsing `https://home.barclays/' as `home' rather than the more relevant `barclays'.
\looseness=-1
% \vspace{-10px}

\observ{Models that rely on brand domain checking heavily depend on the structure of URLs, the URL parsing method, and the comparison against the maintained second-level domain}

\vspace{-10px}
\subsection{Case Study of Failures} \label{subsec:casestudy} 
\texttt{PhishIntention} leverages OCR to incorporate textual information, outperforming \texttt{Phishpedia} on textual logos. For example,~\texttt{PhishIntention} correctly identifies the brand of~\autoref{fig:eg1}, while \texttt{Phishpedia} misclassifies it as `timeweb,' highlighting the crucial role of OCR in logo character recognition.
We further check the manipulated `YouTube' logos in \autoref{tab:manipulation-vicomp-w-img} against brand reference lists. \texttt{Phishpedia} shows lower similarity scores (0.5--0.6) for `Elimination,' `Color Replace,' and `Integration,' while other manipulations are around 0.9. \texttt{PhishIntention} considers all as benign due to the absent CRP in HTML. 
\texttt{PhishZoo} successfully identifies candidate keywords with varying similarity scores across manipulations. \texttt{Involution} mostly fails to recognize the brand (similarity around 0.57), except misidentifying the `Case Conversion' example as `AOL.' \texttt{VisualPhishNet} misclassifies all examples as other brands (scores 1.1--1.3), while \texttt{EMD} predicts `Airbnb' (distance 0.96). The results indicate the difficulty in setting appropriate similarity score thresholds for each solution.

\observ{Textual information of visual elements and appropriate similarity thresholds significantly impact performance. Future models should integrate advanced OCR, human-centric similarity metrics, and multi-modal analysis combining visual and contextual information}

% \begin{figure}[!t]
%     \centering
%     \begin{subfigure}{0.15\textwidth}
%     \includegraphics[width=\textwidth,height=2em]{case_study/rackspace_targetlist.png}
%     \caption{Rackspace brand target list example}
%     \end{subfigure}\hfill
%     \begin{subfigure}{0.15\textwidth}
%     \includegraphics[width=\textwidth,height=2.5em]{case_study/rackspace_benign.png}
%     \caption{Recent searched Rackspace logo}
%     \end{subfigure}\hfill
%     \begin{subfigure}{0.15\textwidth}
%     \includegraphics[width=\textwidth,height=2em]{case_study/timeweb.png}
%     \caption{Misidentified target logo}
%     \end{subfigure}
%     % \vspace{-5px}
%     \caption{Text Logo Case (Original logo and benign URL).}
%     \label{fig:eg1}
%     % \vspace{-10px}
% \end{figure}

\vspace{-15px}
\section{Discussion}
\label{sec:discussion} 

\PPNS{Recommendations}
We propose several key improvements focused on comprehensive threat detection and resilience against manipulation attacks. First, integrating advanced text recognition capabilities through OCR-aided deep learning models or online search verification would significantly strengthen brand identification accuracy. This enhancement would address current limitations in systems (\texttt{Phishpedia}) that struggle with semantic variations in phishing attempts.

Second, detection systems must strengthen their defenses through comprehensive adversarial training that incorporates manipulated logos and visual elements. This can be achieved by systematically exposing machine learning models to real-world manipulation patterns during the training phase. Organizations should implement sophisticated data augmentation techniques that account for common visual modifications, including scaling transformations, color adjustments, and other alterations that may be used to evade detection.

Third, we recommend implementing a holistic, multi-modal detection approach. It should integrate analysis across multiple dimensions: examining logo characteristics, evaluating webpage structural elements, assessing textual content authenticity, and analyzing additional visual indicators.

Finally, we recommend using preprocessing and normalization techniques, including image scaling and denoising, before visual similarity analysis. These methods can reduce the efficacy of adversarial manipulations and provide additional layers of defense against sophisticated phishing tactics.
% \looseness=-1

\PP{Limitations}
Our work has a few key limitations. First, the lack of a user study limits the assessment of our manipulation methods' effectiveness in real-world scenarios, as users may readily recognize manipulated logo images. To address this, we perform manual verification with 500 randomly generated images to ensure our manipulations are not easily recognizable. However, a comprehensive user study would be valuable for gathering insights into the perceptibility and deceptiveness of adversarial manipulations. 

Second, our analysis is limited to logo manipulations and does not consider other webpage elements that could be targeted. Expanding the scope could provide a more comprehensive understanding of potential attack vectors. Third, our evaluation is conducted only on models and datasets with publicly available source code and data. Despite these limitations, we highlight the potential vulnerabilities of visual similarity-based anti-phishing systems and the need for robust defense mechanisms against adversarial visual manipulations.
\vspace{-10px}
\section{Related Work}
\PP{Visual Similarity-based Detection}
Visual similarity-based phishing detection systems compare suspicious websites against legitimate ones to identify threats. While Panum~\etal~\cite{Thomas2020Phishing} and Abuadbba~\etal~\cite{abuadbba2022web} examined detector robustness and evolving phishing trends, their evaluations lacked comprehensive real-world testing. Literature reviews by Zieni~\etal~\cite{Zieni2023PhishingSurvey} and Hou~\etal~\cite{Hou2023LogoSurvey} documented detection techniques but provided no comparative performance analyses.
Our work presents the first comprehensive evaluation of visual similarity-based phishing detectors in controlled environments with consistent brand knowledge across models. We expand upon Liu~\etal~\cite{Liu2022phishintention}'s \texttt{PhishIntention} evaluation by testing detector performance on extensive APWG phishing data and incorporating explainable AI techniques based on LIME~\cite{lime}, similar to Charmet~\etal~\cite{Charmet2024Trans}'s approach, to understand brand impersonation detection.

\PP{Evaluation of Robustness against Adversarial Manipulation} Adversarial attacks use carefully crafted perturbations to manipulate machine learning model predictions~\cite{Goodfellow2015FGSM, madry2018towards}. While Lee~\etal~\cite{lee2023attacking} demonstrated how perturbation vectors could bypass phishing detectors, their analysis of visual impact was limited (\ie, perturbations on the logo components). Similarly, Ying~\etal~\cite{Yuan2024WWW}'s user studies on webpage modifications overlooked the resilience of logo-based detection systems. Hao~\etal~\cite{Look2024Hao} also worked on logo manipulation, which focused primarily on style and font modifications, our study examines detector robustness against a comprehensive range of techniques, including 14 visible manipulations and 5 adversarial perturbations observed in real phishing attacks.
% \looseness=-1

\vspace{-10px}
\section{Conclusion} \label{sec:conclution}
In this comprehensive evaluation of seven visual similarity-based anti-phishing models across 451k real-world phishing websites, we identified substantial performance disparities between controlled testing environments and real-world applications. Our analysis exposed critical weaknesses that could be exploited through adversarial visual manipulations. To strengthen these systems, we recommend integrating text recognition with visual analysis, implementing data augmentation with adversarial examples, adopting a multi-cue ensemble approach, and utilizing preprocessing techniques such as scaling and denoising. These enhancements are essential for developing more robust and reliable phishing detection systems capable of addressing real-world threats.

\vspace{-10px}
\section{Acknowledgments}
We sincerely thank the anonymous shepherd and all the reviewers for their constructive comments and suggestions. This
work is supported in part by the NSF (2210137 and 2335798), a seed grant from the AI Tennessee Initiative at the University of Tennessee Knoxville, Science Alliance’s StART program,  gifts from Google exploreCSR, and IITP grants from South Korean government (RS-2024-00439762 and RS-2024-00419073).
Any opinions, findings, and conclusions or recommendations expressed in this material are those of the authors and do not necessarily reflect the views of the sponsors.

% \vspace{-10px}
\section{Ethics Consideration} \label{sec:ethics}
Our research involving the APWG \texttt{eCX} dataset focused solely on reported phishing websites, ensuring that no benign, legitimate sites were affected. Importantly, no personal data from users or phishing websites was collected or used in our study. To maintain ethical standards, we share only selected failed phishing examples, providing HTML and screenshot data without revealing URLs from the APWG dataset. 

Our research utilizes exclusively open-source models, and the techniques we examine may be employed by malicious actors. By openly sharing our source code and findings, we aim to strengthen cybersecurity defenses against phishing attacks. We believe the security benefits of transparency -- enabling defenders to better understand and counter these threats -- outweigh the potential risks, particularly since attackers already know these methods. This open approach aligns with our commitment to advancing collective cybersecurity capabilities.

\vspace{-10px}
\section{Open Science}
To facilitate reproducibility and accelerate scientific progress (\ie, strengthening collective efforts in combating phishing attacks), we publicly share: (1) our collected datasets, (2) code, and (3) retrained models. The resources are available on our website (\url{https://moa-lab.net/evaluation-visual-similarity-based-phishing-detection-models/}) or Zenodo (\url{https://zenodo.org/records/14668190}).

\PP{Collected Datasets}
We publicly share the 451,514 real-world phishing data that our web crawler collected.
This dataset includes both HTML source files and visual screenshots of phishing websites. Due to licensing agreements with the Anti-Phishing Working Group (APWG), the original phishing URLs are withheld from public sharing. Moreover, we share our extended reference list ($\boldsymbol{R_{ext}}$) that is used for our evaluation (see~\autoref{sec:experiment:overview}). Furthermore, we share a general benign dataset that covers 100 Tranco domains.

\PP{Manipulated Phishing Screenshots}
We publicly share 7,223 manipulated screenshots, including 110 original and all manipulated screenshots.

\PP{Failed Sampled Screenshots and HTML}
We publicly share 6,000 failed, sampled screenshots and HTML with detailed CSV files documenting model-specific failure cases for phishing detection.

\PP{Code}
We publicly share all our code for collecting datasets and evaluating models under an open-source license: (1) testing code with an open-source license, (2) preprocessing codes for clustering, (3) web crawler source code, and (4)
perturbation-based attacking code.

\PP{Retrained Models}
We have retrained three models (PhishIntension, Phishpedia, and Involution) for evaluation. We publicly share our retrained models  as they are clearly under MIT or CC0-1.0 license, which explicitly permits model modification and redistribution. 

%-------------------------------------------------------------------------------
% \bibliographystyle{plain}
% % \bibliographystyle{unsrturl}
% \bibliography{ref}

\appendix

\section{Appendix} \label{sec:appendix}

% Resources of the paper are available at Zenodo (\url{https://doi.org/10.5281/zenodo.14668190}). 
Specific references to the corresponding contents are specified in the following.

\vspace{-5px}
\subsection{Model Summary} ~\label{subsec:model_summary} 
We conduct a comprehensive literature review of top conferences and highly cited papers from 2005 to 2023 to identify popular visual similarity-based models for phishing detection, as summarized in \autoref{tab:referece_based_model}.
% We selected six promising models to retrain and evaluate from this search in our study, as summarized in~\autoref{tab:referece_based_model}. 
The gray sharing in the table indicates the seven models selected for re-training and evaluation.
%
% The selected models span from 2005 to 2022 and employ diverse techniques for assessing websites' visual similarity. 
Liu~\etal~\cite{Liu2005WWW} compare features like layout, colors, fonts, and image placement, while EMD~\cite{Fu2006EMD} uses Earth Mover's Distance to measure visual similarity. Medvet~\etal~\cite{Medvet2008SP}, CCH~\cite{Chen2009IEEEIC}, and Goldphish~\cite{dunlop2010goldphish} analyze discriminative key points, employ image hashing techniques, and leverage classifiers for phishing detection, respectively. More recent approaches like Phishpedia~\cite{Lin2021phishpedia} combine text and visual content analysis, while OpenGlue~\cite{viniavskyi2022openglue}, Bernabeu~\etal~\cite{bernabeu2022multi}, OSLD~\cite{bastan2019large}, Bhurtel~\etal~\cite{Bhurtel2022INFOCOM}, and SeeTek~\cite{Li2022WACV} explore advanced deep learning techniques for image retrieval, logo recognition, and visual similarity assessment. In particular, the use of deep learning (DL) techniques since 2019 shows a growing trend for visual similarity-based phishing detection.

\vspace{-5px}
\subsection{Selected Models} ~\label{sec:select_model}
Based on the candidate papers, we carefully selected seven models in \autoref{tab:compared_model} with diverse architectures, input types, and detection methods to compare visual similarity-based phishing detection approaches comprehensively. Particularly, four of them take screenshots, URLs, and HTML as input, while three of them take screenshots as input.

% Our selection includes two deep learning models (\texttt{PhishIntention}~\cite{Liu2022phishintention} and \texttt{Phishpedia}~\cite{Lin2021phishpedia}), which are specifically designed to identify phishing target brands through logo components. We also selected a deep-learning model that detects phishing using screenshots (\texttt{VisualPhishNet}~\cite{Abdelnabi2020visualphishnet}), and a deep-learning model (\texttt{Involution}~\cite{viniavskyi2022openglue}) not explicitly targeted to detect phishing, and two traditional machine learning approaches, \texttt{EMD} (Earth Mover’s Distance)~\cite{Fu2006EMD} for screenshot-based phishing detection and \texttt{PhishZoo}~\cite{Afroz2011PhishZoo} for logo-based detection.
% % Detailed model information can refer to ~\autoref{tab:compared_model}.
% %
% By including models with diverse architectures, input types, and detection methods, we aim to provide insights into the strengths and limitations of various visual similarity-based phishing detection techniques. 
% % The detailed information for each selected model can be found in \autoref{tab:compared_model}.

\vspace{-5px}
\subsection{FLOPs and Parameters Performance}~\label{sec:flop_detail}
We compare the FLOPs and parameters for key model components in~\autoref{tab:flop_parameter}. \texttt{DynaPhish}, \texttt{PhishIntention}, and \texttt{Phishpedia} have similar model structures, resulting in close parameters and FLOPs for detecting logos and siamese modules. We exclude the CRP locator and web interaction parts of \texttt{PhishIntention} and \texttt{DynaPhish},
as our URLs may not be alive now. Furthermore, the online search function of \texttt{DynaPhish} is not included in the calculation. \texttt{Involution} employs the same module with \texttt{Phishpedia} for logo cropping and thus shares the parameter size. \texttt{EMD} and \texttt{PhishZoo} are not taken into consideration because they do not use neural networks.
\looseness=-1

% \begin{figure*}[!t]
% \centering
% \begin{subfigure}{0.15\textwidth}
% \includegraphics[width=\linewidth,height=3em]{fgm.png}
% \caption{FGSM}
% \end{subfigure}\hfill
% \begin{subfigure}{0.15\textwidth}
% \includegraphics[width=\linewidth,height=3em]{pgd.png}
% \caption{PGD}
% \end{subfigure}\hfill
% \begin{subfigure}{0.15\textwidth}
% \includegraphics[width=\linewidth,height=3em]{cw.png}
% \caption{CW}
% \end{subfigure}
% \vspace{1em}
% \begin{subfigure}{0.15\textwidth}
% \includegraphics[width=\linewidth,height=3em]{vit_crop_logo.png}
% \caption{ViT}
% \end{subfigure}\hfill
% \begin{subfigure}{0.15\textwidth}
% \includegraphics[width=\linewidth,height=3em]{swin_crop_logo.png}
% \caption{Swin}
% \end{subfigure}\hfill
% \begin{subfigure}{0.15\textwidth}
% \includegraphics[width=\linewidth,height=3em]{srnet.png}
% \caption{SRNet}
% \end{subfigure}
% \caption{Perturbated Manipulation Samples Cropped by Faster-RCNN.}~\label{fig:craft_perturbated}
% \end{figure*}

\begin{figure}[!t]
\centering
\begin{subfigure}{0.15\textwidth}
\includegraphics[width=\linewidth,height=3em]{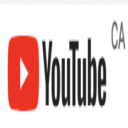}
\caption{FGSM}
\end{subfigure}\hfill
\begin{subfigure}{0.15\textwidth}
\includegraphics[width=\linewidth,height=3em]{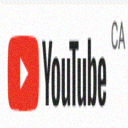}
\caption{PGD}
\end{subfigure} \hfill
\begin{subfigure}{0.15\textwidth}
\includegraphics[width=\linewidth,height=3em]{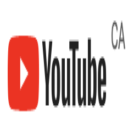}
\caption{CW}
\end{subfigure}
% \vspace{1em}
\begin{subfigure}{0.15\textwidth}
\includegraphics[width=\linewidth,height=3em]{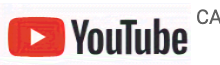}
\caption{ViT}
\end{subfigure}\hfill
\begin{subfigure}{0.15\textwidth}
\includegraphics[width=\linewidth,height=3em]{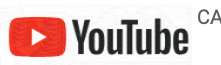}
\caption{Swin}
\end{subfigure}\hfill
\begin{subfigure}{0.15\textwidth}
\includegraphics[width=\linewidth,height=3em]{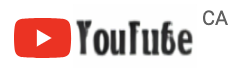}
\caption{SRNet}
\end{subfigure}
% \vspace{-10px}
\caption{Perturbated Logos Cropped by Faster-RCNN.}~\label{fig:craft_perturbated}
\vspace{-15px}
\end{figure}
\begin{figure}[t]
    \centering
    \begin{subfigure}{0.15\textwidth}
    \includegraphics[width=\textwidth,height=2em]{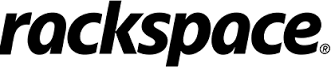}
    \caption{Rackspace brand target list example}
    \end{subfigure}\hfill
    \begin{subfigure}{0.15\textwidth}
    \includegraphics[width=\textwidth,height=2.5em]{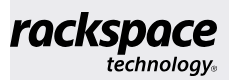}
    \caption{Recent searched Rackspace logo}
    \end{subfigure}\hfill
    \begin{subfigure}{0.15\textwidth}
    \includegraphics[width=\textwidth,height=2em]{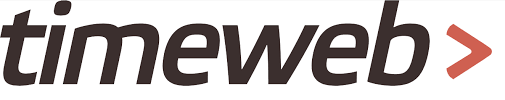}
    \caption{Misidentified target logo}
    \end{subfigure}
    % \vspace{-5px}
    \caption{Text Logo Case (Original logo and benign URL).}
    \label{fig:eg1}
    \vspace{-15px}
\end{figure}
\begin{table*}[!t]
    \renewcommand{\arraystretch}{1.2}
    \setlength{\tabcolsep}{3pt}
    \normalsize
    \caption{List of Visual Similarity-based Models (Y/N: open source code or not, Y*: reproduced by non-original authors).}
    
    % \KK{The column ``model name'' seems the title of papers. As the paper titles are available in the reference, how about adding the model's name and brief description? I suggest adding one more column for a model architecture. Also, the columns that use a deep learning model can be informative.}
    \label{tab:referece_based_model}
    \centering
    \resizebox{1\linewidth}{!}{
    % \begin{tabular}{lp{.147\textwidth}p{.35\textwidth}clr}
    \begin{NiceTabular}{llllll}
        \toprule
        \textbf{Year} & \multicolumn{1}{c}{\textbf{Model}} & \multicolumn{1}{c}{\textbf{Description}} & \multicolumn{1}{c}{\textbf{DL}} & \textbf{Code} & \textbf{Data Source}\\ 
        \midrule
        2005 & Liu~\etal~\cite{Liu2005WWW} & Compares features like layout, colors, fonts, and image placement of webpages for phishing detection & N & N & N\\ 
        \rowcolor{lightgray!30} 
        2006 & EMD~\cite{Fu2006EMD} & Uses Earth Mover's Distance to assess visual similarity of webpages for phishing detection & N &  Y$^*$~\cite{lindsey932020online} & N \\
        2008 & Medvet~\etal~\cite{Medvet2008SP} & Relies on visual similarity, comparing features like layout, colors, and overall webpage appearance, to detect phishing & N & N & PhishTank, Alexa\\ % NO put in the venue.
        2009 & CCH~\cite{Chen2009IEEEIC} & Employs discriminative keypoint features to distinguish phishing websites based on visual cues & N & N & N\\
        2010 & Goldphish~\cite{dunlop2010goldphish} & Analyzes images for phishing detection, possibly using techniques like image recognition or text extraction from images & N & N & PhishTank\\
        2011 & Zhang~\etal~\cite{Zhang2011IEEETrans} & Combines textual and visual content analysis with a Bayesian approach for phishing detection & Y & N & N \\
        2011 & msDT~\cite{kalantidis2011scalable} & Introduces a method for logo recognition (not phishing specific) based on triangulation & N & N & Flickr\\
        \rowcolor{lightgray!30}
        2011 & PhishZoo~\cite{Afroz2011PhishZoo} & Analyzes visual appearance of webpages, likely using techniques to compare layout, colors, fonts, and images & N & Y$^*$~\cite{lindsey932020online} & PhishTank, Alexa\\
        
        2013 & Chang~\etal~\cite{Chang2013ICITCS} & Focuses on website identity recognition, using techniques like domain name analysis or website structure comparison & N & N & PhishTank, Alexa\\ % also use google, but can be taken as benign one
        2013 & Romberg~\etal~\cite{Romberg2013ICMR} & Proposes bundle min-hashing for logo recognition & N & Y & Flickr\\
        2015 & FaceNet~\cite{schroff2015facenet} & Deep learning architecture for face recognition (not directly related to phishing) & Y & Y & LFW, YoutubeDB\\
        2015 & Rao~\etal~\cite{Rao2015IEEECSNT} & Presents a computer vision technique for phishing detection using visual similarity & N & N & PhishTank \\
        2015 & LOGO-Net~\cite{hoi2015logo} & Leverages deep learning for logo detection (not directly related to phishing) & Y & N & N\\
        2016 & Bozkir~\etal~\cite{Bozkir2016ISDFS} & Uses HOG descriptors for feature extraction to potentially compare webpages for phishing detection & N & N & N\\
        2017 & DeltaPhish~\cite{corona2017deltaphish} & Compares the static features of HTML and visual appearance of the potential phishing pages against compromised websites & N & N & PhishTank\\
        2017 & Haruta~\etal~\cite{Haruta2017IEEEGCC} & Combines image and CSS analysis with a target website finder for phishing detection & N & N & Alexa\\
        2019 & Sharma~\etal~\cite{sharma2019retrieving} & Deep learning approach for image retrieval (adaptable to phishing) & Y & Y & N\\
        2020 & CSQ~\cite{Yuan2020CVPR} & Deep learning method for image/video retrieval (adaptable to phishing) & Y & Y & ImageNet\\
        \rowcolor{lightgray!30}
        2020 & VisualPhishNet~\cite{Abdelnabi2020visualphishnet} & Proposes zero-day phishing website detection based on visual similarity & Y & Y & N\\
        \rowcolor{lightgray!30}
        2021 & Involution~\cite{Li2021CVPR} & Inverting the inherence of convolution for visual recognition & Y & Y & Cityscapes\\
        2021 & Dooremaal~\etal~\cite{Dooremaal2021ARES} &  Combining text and visual features to improve the identification of cloned webpages for early phishing detection &Y & N & Y\\
        \rowcolor{lightgray!30}
        2021 & Phishpedia~\cite{Lin2021phishpedia} & Employs a hybrid deep learning approach for visual phishing detection & Y & Y & Phishpedia\\
        \rowcolor{lightgray!30}
        2022 & PhishIntention~\cite{Liu2022phishintention} & Uses deep learning to analyze webpage appearance and dynamics for inferring phishing intention & Y & Y & PhishiIntention\\
        2022 & OpenGlue~\cite{viniavskyi2022openglue} & Open-source deep learning pipeline for image matching (not directly related to phishing) & Y & Y & MegaDepth\\
        2022 & Bernabeu~\etal~\cite{bernabeu2022multi} & Leverages deep learning for multi-label logo recognition (not directly related to phishing) & Y & N & METU \\
        2022 & OSLD~\cite{bastan2019large} & Deep learning approach for large-scale logo detection (not directly related to phishing) & Y & N & OSLD\\
        %  dataset Y, code N (he said N on github issue)\\
        2022 & Bhurtel~\etal~\cite{Bhurtel2022INFOCOM} & Relies on machine learning with a Siamese network for logo recognition for phishing detection & Y & N & LogoSENSE\\
        2022 & SeeTek~\cite{Li2022WACV} & Deep learning for large-scale logo recognition with text integration (not directly related to phishing) & Y & N & PL8K\\
        \rowcolor{lightgray!30}
        2023 & DynaPhish~\cite{Dynaphish2023Liu} & Using deep learning approaches to analyze webpages and Google search to identify brand intention  & Y & Y & DynaPhish\\
        \midrule
        % 2013 & GoldPhish: Using Images for Content-Based Phishing Analysis~\cite{Dunlop2013GoldPhish} \\ ThirdParty
        % 2013 & BaitAlarm: Detecting Phishing Sites Using Similarity in Fundamental Visual Features~\cite{Mao2013BaitAlarm} \\ %CSS
        % 2017 & Phishing-Alarm: Robust and Efficient Phishing Detection via Page Component Similarity~\cite{Mao2017PhishAlarm} \\
        % 2018 & \makecell[l]{Phish-IRIS: A New Approach for Vision Based Brand Prediction of Phishing Web Pages via Compact Visual Descriptors~\cite{Dalgic2018PhishIRIS}} \\
        \multicolumn{5}{l}{Where deeper shades of \begingroup\fboxsep=0pt\colorbox{lightgray!30}{\reducedstrut \phantom{000}}\endgroup ~ indicate the seven models that we select for retraining and evaluation.}\\
    \end{NiceTabular}
    }
    \vspace{-15px}
\end{table*}

\begin{table}[!t]
    \renewcommand{\arraystretch}{1}
     \normalsize
    \caption{Components' FLOP and Parameters Performance.}
    \label{tab:flop_parameter}
    \centering
    \resizebox{\linewidth}{!}{
    \begin{NiceTabular}{lcccccc}
        \toprule
        \multirow{2}{*}{\textbf{Model}} & \multicolumn{4}{c}{\textbf{Parameters/FLOPs}} \\ 
        \cmidrule{2-7}
        & \textbf{Detect Logo} & \textbf{Siamese} & \textbf{CRP Classifier} & \textbf{Others} & \textbf{Total} \\ 
        \midrule
        DynaPhish  & 41.32M/203G & 24.10M/1.35G & 23.50M/11.31G & --- & 88.92M/215.66G\\
        PhishIntention  & 41.32M/203G & 24.10M/1.35G & 23.50M/11.31G & --- &88.92M/215.66G\\
        Phishpedia  &  41.30M/211G & 24.10M/1.35G & --- & --- & 65.40M/212.35G\\
        Involution  &  41.30M/211G & --- &--- & 12.01M/1.67G & 53.04M/212.67G&\\
        VisualPhishNet  &---  &---&---& 21.27M/92.49G & 21.27M/92.49G\\
        \bottomrule
    \end{NiceTabular}
    }
    \vspace{-10px}
\end{table}

\begin{table}[!t]
    \caption{Failure Categorization in Our Dataset.}
    \label{tab:pass_screenshot_category}
    \centering
    % \normalsize
    \resizebox{1\linewidth}{!}{
    \begin{NiceTabular}{llll}
        \toprule
        & \textbf{ID} & \textbf{Name} & \textbf{Description} \\ \midrule
        \multirow{22}{*}{Logo} 
        & L1 & Similar & Similar to the reference list \\
        & L2 & Elimination & Screenshots delete logos \\
        & L3 & BrokenImage & Logo images are damaged \\
        & L4 & ColorReplace & Different colors of logos \\
        & L5 & LogoBackground & Different backgrounds of logos \\
        & L6 & Integration & Logos are combined with other logos \\
        & L7 & Re-position & Logos appear in different locations on the screenshot \\ 
        & L8 & Outdated & Logos are not in the reference list \\
        & L9 & CaseConversion & Changing the case of textual logos \\
        & L10 & TextAsLogo & Type text as the logo \\
        & L11 & Scaling & Enlarge or shrink logos \\ 
        & L12 & Resizing & Logos' height-to-width are changed \\
        & L13 & FontReplace & Changing the textual font of logos \\
        & L14 & Omission & Only partial logos are used \\
        & L15 & Shape & Logo with different shapes, like square, rectangular \\
        & L16 & LogoAddText & Add text close to the logos \\
        & L17 & Replacement & Screenshots replace logos with other logos\\ 
        & L18 & Rotation & Logos are rotated in some angles\\
        & L19 & Flipping & Flipping logo by vertical or horizontal\\ 
        & L20 & Blurring & The logo or screenshot is blurred \\
        & L21 & CraftLogo & Craft logos based on different information\\ 
        & L22 & Language & Change the textual logos language \\ \midrule
        \multirow{4}{*}{Popup} 
        & P1 & LoginPopup & Login forms pop-up on the screenshot \\
        & P2 & AdPopup & Advertisements pop-up on the screenshot \\ 
        & P3 & CookiePopup& The cookie pop up on the blurred screenshot\\ 
        & P4 & OtherPopup & Alert, remind, location, etc. \\ \midrule
        \multirow{5}{*}{Login} 
        & F1 & LoginForm & Change login form text (text, color, language, fonts)\\
        & F2 & Button & Change button color, shape, location, text, etc. \\ 
        & F3 & NewForm & Design a new form \\ 
        & F4 & ThirdParty & Use other websites as login methods \\ 
        & F5 & QR & Login by scanning the QR code \\ \midrule
        \multirow{3}{*}{Others}
        & O1 & ImageAddText & Add text on the screenshot not close to logo areas\\
        & O2 & Blocked & The image is blocked, only left text \\
        & O3 & ImageBackground & Different backgrounds of screenshots \\
        \bottomrule
    \end{NiceTabular}}
    \vspace{-10px}
\end{table}

\vspace{-5px}
\subsection{Failure Examples Categorization}~\label{sec:pass_example_categorization}
To better understand the limitations of visual similarity-based phishing detection models, we analyzed the failure cases observed during our real-world evaluations in~\autoref{sec:results}. We categorized these failure examples into four main categories: logo, popup, login form, and other related issues, as summarized in \autoref{tab:pass_screenshot_category}.
Logo-related issues (L1-L22) encompass various manipulations and alterations to the logo, such as elimination and color replacement. These issues highlight models' challenges in accurately identifying and comparing logos under diverse visual variations.
Popup-related issues (P1-P4) pose a significant threat. They involve the presence of popups, advertisements, cookies, alerts, and other overlays on the screenshot. These elements can obstruct or confuse the visual analysis of the web pages, potentially leading to misclassifications by the phishing detection models and, consequently, to successful phishing attacks.
Login form-related issues (F1-F5) include changes to the login form's text, color, language, font, and other website login forms as a phishing tactic. These variations in the login form's appearance and design can make it difficult for models to identify phishing attempts based on visual similarity alone accurately. Other manipulations (O1-O3) include adding extra text on the screenshot and some pages that are blocked.

\vspace{-5px}
\subsection{Perturbated and SRNet}~\label{sec:perturbated_example}
The perturbed and SRNet logo samples cropped from screenshots by Faster-RCNN are shown in \autoref{fig:craft_perturbated}. Three white-box attacks, two black-box attacks, and SRNet methods are summarized as follows:\\
% \PP{Fast Gradient Sign Method (FGSM)}~\cite{Goodfellow2015FGSM}: A white-box attack that adds perturbations to the logo based on the gradients of the target model, calculated in a single step.\\
% \PP{Projected Gradient Descent (PGD)}~\cite{madry2018towards}: An iterative version of the FGM attack, which applies the perturbations multiple times to create a more effective adversarial example.\\
% \PP{Carlini \& Wagner (CW)}~\cite{Carlini2017CW}: A strong white-box attack that optimizes the perturbations to minimize the detection confidence of the target model.\\
% \PP{Vision Transformer (ViT)}~\cite{dosovitskiy2021ViT}: A transformer-based architecture for image classification that splits an image into patches and processes them using self-attention mechanisms.\\
% \PP{Swin Transformer (Swin)}~\cite{liu2021swin}: A hierarchical vision transformer that uses shifted window attention to capture both local and global dependencies in an image efficiently.\\
% \PP{Style Retention Network (SRNet)}~\cite{WuEdit2019}: A generative adversarial network (GAN) that aims to transfer the style of one image to another while preserving the content of the target image.
\PP{Fast Gradient Sign Method (FGSM)}~\cite{Goodfellow2015FGSM}: A white-box attack that perturbs the data in a single step by using an imperceptibly small vector, elements are equal to the sign of the gradient of the cost function with respect to the input.\\
\PP{Projected Gradient Descent (PGD)}~\cite{madry2018towards}: An iterative version of the FGM attack with a random start, which applies the perturbations multiple times to create a more effective adversarial example.\\
\PP{Carlini \& Wagner (CW)}~\cite{Carlini2017CW}: A strong white-box attack that optimizes the perturbations to minimize the detection confidence of the target model. Constructing three new attacks for the three distance metrics.\\
\cite{lee2023attacking}-\textbf{ViT} and \cite{lee2023attacking}-\textbf{Swin.} Black-box attacks that utilize generative adversarial perturbations to develop adversarial logos. Taking the trained Vision Transformer (ViT)~\cite{dosovitskiy2021ViT} and Swin Transformer~\cite{liu2021swin} models as Discriminators and a Deep Residual Network with six residual blocks (ResNet-6) as the foundational architecture of the Generator.\\
\PP{Style Retention Network (SRNet)}~\cite{WuEdit2019}: A generative adversarial network (GAN) that aims to transfer the style of images to another while preserving the content.

\vspace{-10px}
\subsection{Examples of Visible Manipulation}~\label{sec:failure_example}
\begin{figure*}[hbt!]
\centering
\begin{subfigure}{0.26\textwidth}
\fbox{\includegraphics[width=\linewidth]{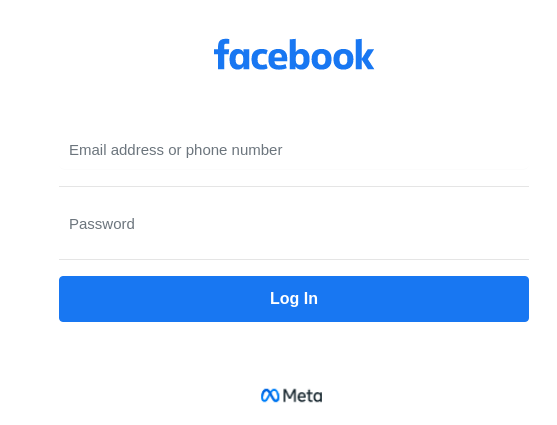}}
\caption{EMD Failed Example}
\label{fig:a}
\end{subfigure}\hfill
\begin{subfigure}{0.25\textwidth}
\fbox{\includegraphics[width=\linewidth]{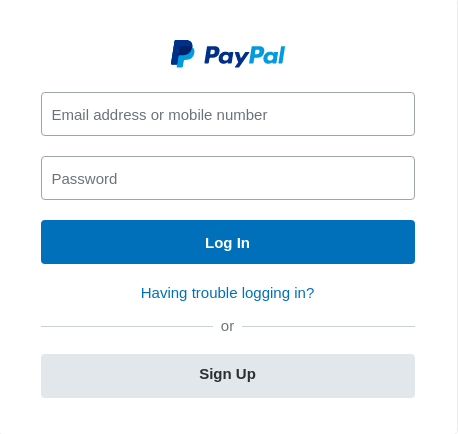}}
\caption{VisualPhishNet Failed Example}
\label{fig:b}
\end{subfigure}\hfill
\begin{subfigure}{0.25\textwidth}
\fbox{\includegraphics[width=\linewidth]{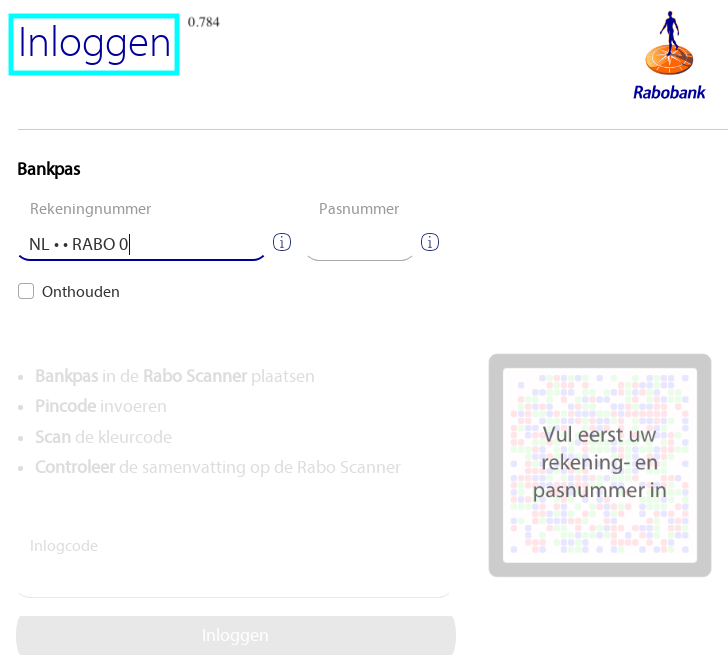}}
\caption{Wrong Logo Area and QR Code}
\label{fig:c}
\end{subfigure}\hfill
\begin{subfigure}{0.27\textwidth}
\hfill\fbox{\includegraphics[width=.9\linewidth]{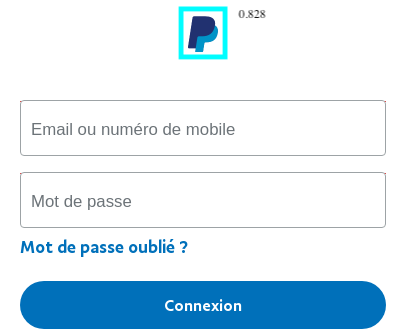}}
\caption{Similar (<Threshold)}
\label{fig:d}
\end{subfigure}\hfill
\begin{subfigure}{0.26\textwidth}
\fbox{\includegraphics[width=\linewidth]{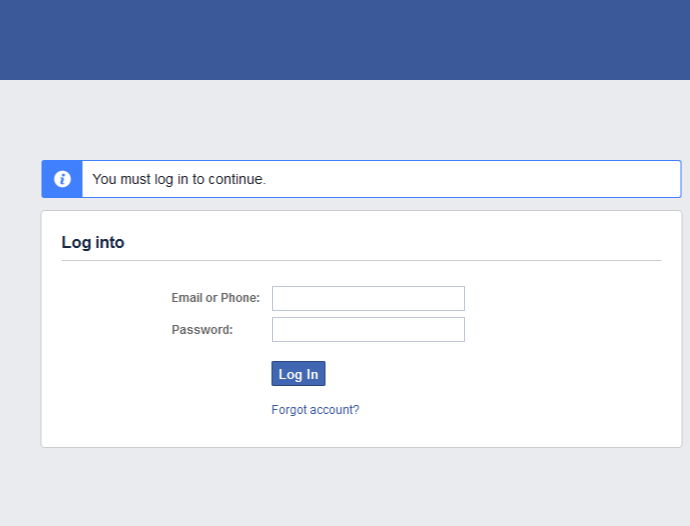}}
\caption{Elimination of Logo}
\label{fig:e}
\end{subfigure}\hfill
\begin{subfigure}{0.27\textwidth}
\fbox{\includegraphics[width=\linewidth]{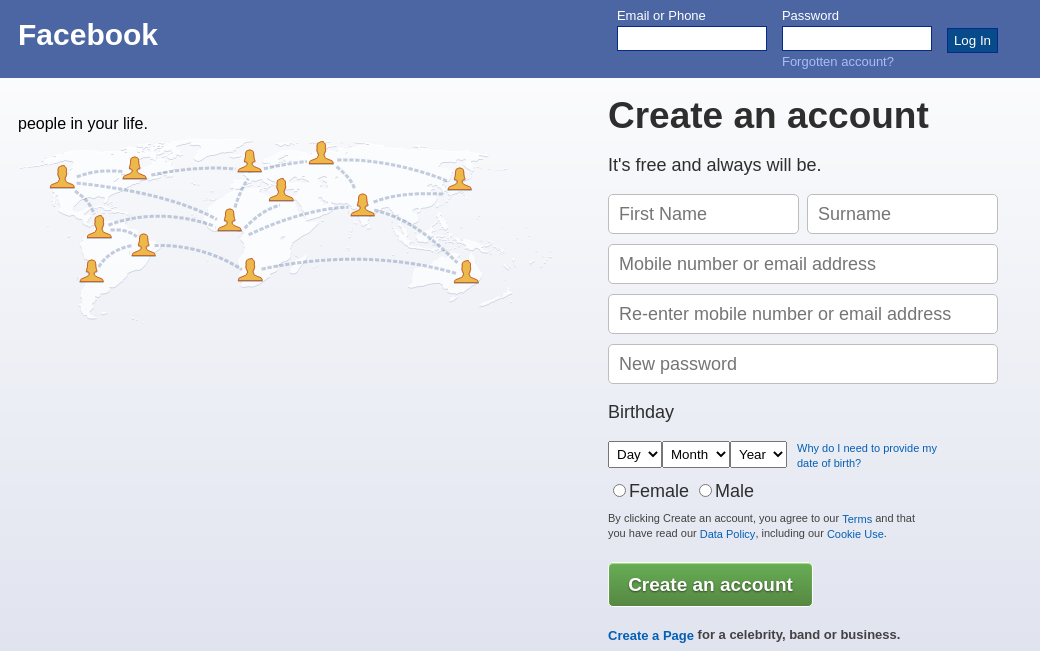}}
\caption{First Letter Upper Case Conversion}
\label{fig:f}
\end{subfigure}\hfill
\begin{subfigure}{0.24\textwidth}
\fbox{\includegraphics[width=\linewidth]{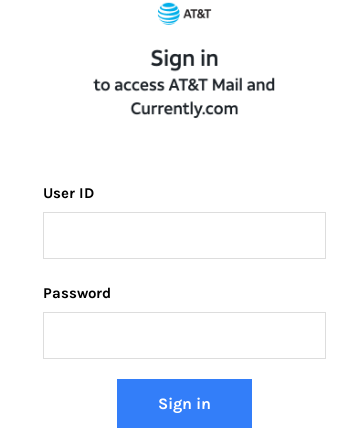}}
\caption{PhishZoo Failed Example}
\label{fig:g}
\end{subfigure}\hfill
\begin{subfigure}{0.26\textwidth}
\hfill\fbox{\includegraphics[width=.9\linewidth]{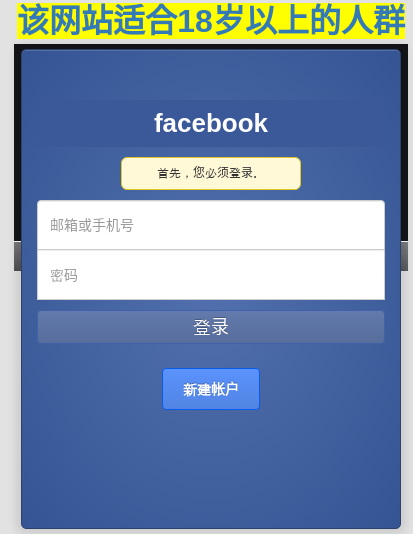}}
\caption{Adding Text}
\label{fig:h}
\end{subfigure}\hfill
\begin{subfigure}{0.26\textwidth}
\fbox{\includegraphics[width=\linewidth]{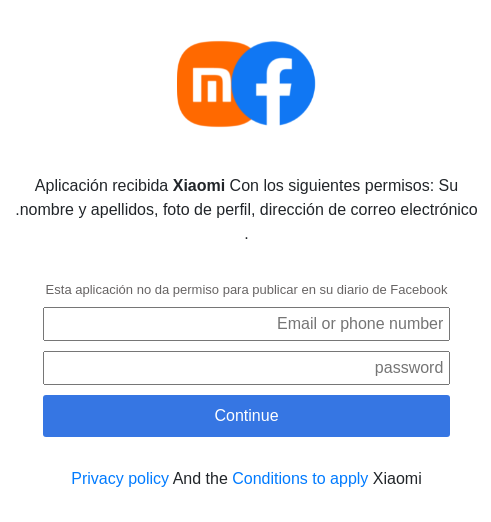}}
\caption{Integration of Logos}
\label{fig:i}
\end{subfigure}\hfill
\begin{subfigure}{0.26\textwidth}
\fbox{\includegraphics[width=\linewidth]{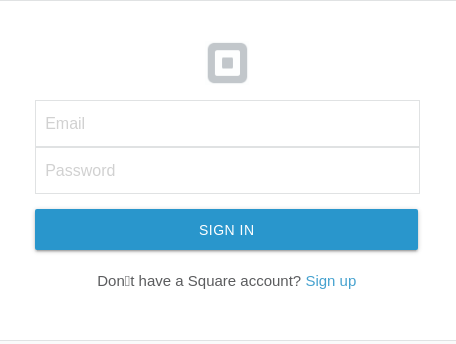}}
\caption{Omission and Color Replacement}
\label{fig:j}
\end{subfigure}\hfill
\begin{subfigure}{0.26\textwidth}
\fbox{\includegraphics[width=\linewidth]{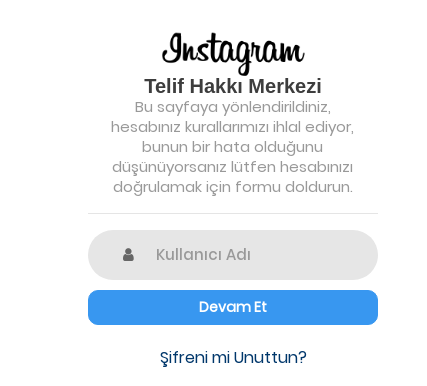}}
\caption{Font Replacement}
\label{fig:k}
\end{subfigure}\hfill
\begin{subfigure}{0.24\textwidth}
\fbox{\includegraphics[width=.9\linewidth]{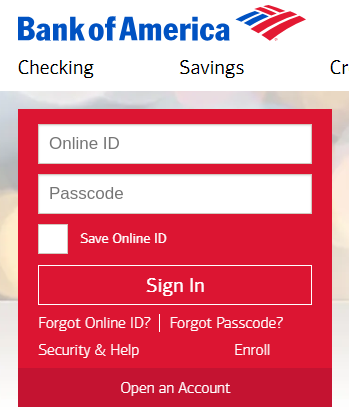}}\hfill
\caption{Case Conversion and Outdated}
\label{fig:l}
\end{subfigure}\hfill
\caption{Examples of Manipulated Samples Found in Our Real-world Phishing Dataset.}~\label{fig:failure_example}
\end{figure*}

Focusing on the logo component, we randomly select the samples that can make models fail. We also selected the samples based on the manipulations used by the adversaries. Detailed examples can refer to ~\autoref{fig:failure_example}.
Additionally, \autoref{fig:eg1} shows an example that \texttt{PhishIntention} correctly identified but \texttt{Phishpedia} failed to recognize.
\looseness=-1

\begin{table*}[!t]
    \setlength{\tabcolsep}{4pt}
    \caption{Description of Used Seven Model Information.} 
    \label{tab:compared_model}
    \vspace{-5px}
    \small
    \resizebox{\textwidth}{!}{
    \begin{NiceTabular}{lllp{.8\linewidth}}
        \toprule
         \multicolumn{1}{c}{\textbf{Model Name}} & \textbf{Training Dataset} & \multicolumn{1}{c}{\textbf{Input}} & \multicolumn{1}{c}{\textbf{Description}} \\ 
        \midrule
        \multirow{1}{*}{EMD}~\cite{Fu2006EMD}  & \multirow{1}{*}{---} & \multirow{1}{*}{S} & Calculate distance by EMD through color and coordinate feature\\ 
        \midrule
        % \multirow{2}{*}{PhishZoo~\cite{Afroz2011PhishZoo}} & --- & \multirow{2}{*}{S, U, H} & Use TF-IDF on URL and HTML for profile matching and use the SIFT feature for image matching \\ 
        \multirow{1}{*}{PhishZoo~\cite{Afroz2011PhishZoo}} & \multirow{1}{*}{---} & \multirow{1}{*}{S, U, H} & Use TF-IDF on URL and HTML for profile matching and  use the SIFT feature for image matching \\
        \midrule
        \multirow{2}{*}{VisualPhishNet~\cite{Abdelnabi2020visualphishnet}} & \multirow{2}{*}{$\boldsymbol{R_{ext}}$}  & \multirow{2}{*}{S} & Use Triplet CNN to learn similarities of the same websites' screenshots and dissimilarities between different websites' screenshots.\\ 
        \midrule
        \multirow{2}{*}{Involution~\cite{Li2021CVPR}} & \multirow{2}{*}{\makecell[l]{Logo2K+, \\ $\boldsymbol{R_{base}}$ or $\boldsymbol{R_{ext}}$}} & \multirow{2}{*}{S} & Use Faster-RCNN to find the logo region, learn logo representations through Involution, and then compare cosine similarity \\ 
        \midrule
        \multirow{2}{*}{Phishpedia~\cite{Lin2021phishpedia}} & \multirow{2}{*}{\makecell[l]{Logo2K+, Benign30K, \\$\boldsymbol{R_{base}}$ or $\boldsymbol{R_{ext}}$}} & \multirow{2}{*}{S, U, H} &  Contains a layout classifier designed to detect and locate the logo region within images, and a Siamese neural network model that analyzes the identified logo to recognize and classify the brand it represents\\ 
        \midrule
        \multirow{3}{*}{PhishIntention~\cite{Liu2022phishintention}} & \multirow{3}{*}{\makecell[l]{Logo2K+, \\Sampled Benign30K, \\$\boldsymbol{R_{base}}$ or $\boldsymbol{R_{ext}}$}} & \multirow{3}{*}{S, U, H} & Contains a layout classier part to find the different components' regions, a CRP classifier to check if the screenshot has CRP, an HTML static classifier to check whether have CRP, a CRP locator to find additional links' CRP, and a Siamese model to recognize the logo's brand\\ 
        \midrule
        \multirow{2}{*}{DynaPhish~\cite{Dynaphish2023Liu}} & \multirow{2}{*}{---} & \multirow{2}{*}{S, U, H} & Based on ~\cite{Liu2022phishintention} and~\cite{Lin2021phishpedia}, it contains a Google search part to check targeted brands and dynamically expand reference lists\\
        \midrule
        \multicolumn{4}{l}{
        *\textbf{Testing Dataset} = APWG Dataset, Manipulating Dataset; **\textbf{Brand Reference List} = Baseline Ref. $\boldsymbol{D_{base}}$, Extended Ref. $\boldsymbol{D_{ext}}$;
        }\\
        \multicolumn{4}{l}{
        \textbf{S} = Screenshot; \textbf{U} = URL; \textbf{H} = HTML.
        }
    \end{NiceTabular}}
   \vspace{-5px}
\end{table*}

\begin{table*}[!t]
    \renewcommand{\arraystretch}{0.8}
    \setlength{\tabcolsep}{2pt}
    \vspace{-5px}
    \caption{Example and Description of Visible Manipulation Methods. 
    }
    \label{tab:manipulation-vicomp-w-img}
    \vspace{-5px}
    \centering
    \small
    \resizebox{0.95\linewidth}{!}{
    \begin{NiceTabular}{p{.12\textwidth} p{.35\textwidth} | P{.12\textwidth} p{.33\textwidth}}
        \toprule
        \multicolumn{1}{c}{\textbf{Example}}& \multicolumn{1}{c}{\textbf{Method Description}} & \multicolumn{1}{c}{\textbf{Example}}& \multicolumn{1}{c}{\textbf{Method Description}} \\ 
        \midrule
        \raisebox{-.7\height}{\includegraphics[clip, scale=0.3, trim={0.3cm 0 1.1cm 0}]{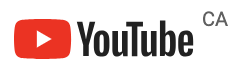}}& \textbf{Original:} This is the original logo cropped from the ``YouTube'' original website.
        &\raisebox{-.7\height}{\includegraphics[clip, scale=0.3, trim={0 0 1.1cm 0}]{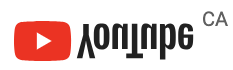}}& \textbf{Flipping:} We flip the logo vertically or horizontally. It differs from ``Rotation,'' where we control the rotation to a small degree.\\
        \midrule
        \raisebox{-1\height}{\includegraphics[clip, scale=0.3, trim={0.5cm 0 1.1cm 0}]{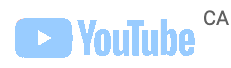}}& %\textbf{Color:} 
        \textbf{Color Replacement:}
        We identify the logo in the screenshot and then replace the color. In this example, we change the original red to blue, but the attacker could use any other predefined color.
        &\raisebox{-1\height}{\includegraphics[clip, scale=0.25, trim={0.5cm 0 0.3cm 0}]{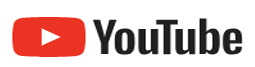}}& %\textbf{Ratio:} 
        \textbf{Resizing:} We randomly modify the height-to-width ratio of the logo. Note that logo resizing does not necessarily maintain the proportion. \\ %In the example shown, the proportions change to the original proportions minus 0.08. \\
        \midrule
        \raisebox{-1.3\height}{\includegraphics[clip, scale=0.3, trim={0 0 1cm 0}]{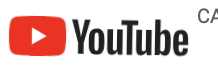}}& \textbf{Rotation:} We rotate the logo in small increments clockwise or counterclockwise, and fill the empty area created by the rotation with the color of the surrounding background. In this example, it is rotated clockwise by one degree.
        &\raisebox{-1.1\height}{\includegraphics[clip, scale=0.3, trim={0 0 0.3cm 0}]{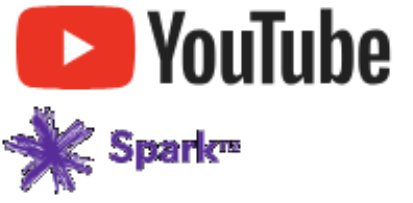}}& %\textbf{Combination:} 
        \textbf{Integration:} We randomly select a second logo from a set of 110 different target brands and place it either above, below, or to the left of the original logo in the screenshot. For example, the ``YouTube'' is combined with ``Spark NZ.’'\\
        \midrule
        \raisebox{-1\height}{\includegraphics[clip, scale=0.3, trim={0.3cm 0 1.2cm 0}]{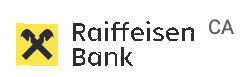}}& \textbf{Replacement:} We replace the original logo with a logo randomly selected from 110 brands. For example, the login form is still ``YouTube,'' but the logo is replaced with ``Raiffeisen Bank.''
        &\raisebox{-1\height}{\includegraphics[clip, scale=0.24, trim={0.3cm 0 0 0}]{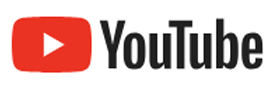}}& \textbf{Scaling:} We scale up the logo, increasing both the length and width to 1.1 times the original size. Then, we place the resized logo in the screenshot of ``Elimination.'' \\ %``Deletion.''\\
        \midrule
        \raisebox{-1.1\height}{\includegraphics[scale=0.2, trim={0.3cm 0 0.3cm 0},width=0.9\textwidth,height=3em]{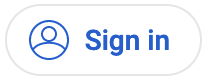}}& %\textbf{Deletion:}
        \textbf{Elimination:} We remove the logo from the screenshot and fill the area with the surrounding background color. The region detector may identify other components (``sign in'') as the logo.
        &\raisebox{-1\height}{\includegraphics[clip, scale=0.3, trim={0.3cm 0 0 0}]{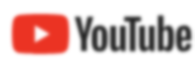}}& \textbf{Blurring:} We add Gaussian blurring with kernel size 9 to the entire screenshot image, including the logo and the background, by the ``OpenCV'' Python package.\\
        \midrule
        \raisebox{-1.3\height}{\includegraphics[clip, scale=0.3, trim={0.3cm 0 0.3cm 0}]{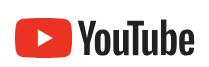}}& %\textbf{Location:} 
        \textbf{Re-position:}
        We move the position of the logo horizontally within the screenshot and fill it with the surrounding background color. The example is cropped from the screenshot when the logo is moved from the top left to the bottom left.
        &\multicolumn{1}{l}{\raisebox{-1.4\height}{\includegraphics[clip, scale=0.3, trim={0 0 0 0}]{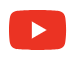}}}& %\textbf{Division:} 
        \textbf{Omission:} We use only one of the elements of the logo (either icon or text) and fill the rest with the surrounding color. For example, we keep the icon and remove the text ``YouTube.''\\
        \midrule
        \raisebox{-1.5\height}{\includegraphics[clip, scale=0.3, trim={0.3cm 0 1.1cm 0}]{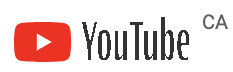}}& %\textbf{Font:} 
        \textbf{Font Replacement:} We use a font identification tool (\url{https://www.myfonts.com/pages/whatthefont}) to find similar fonts. Then, we generate text in those fonts and replace the original logo. We also use the SRNet~\cite{WuEdit2019} to generate text logos while keeping the background context, font style, and color.
        &\raisebox{-1.5\height}{\includegraphics[clip, scale=0.3, trim={0.3cm 0 1.1cm 0}]{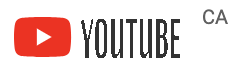}}& %\textbf{Case:} 
        \textbf{Case Conversion:} We find a font that looks similar to the text logo and then change the capitalization of the text to make all letters capitalized, all letters lowercase, or just the first letter capitalized. For example, ``YouTube'' is transformed into ``YOUTUBE.''\\
        \bottomrule
    \end{NiceTabular}
    }
\vspace{-10px}
\end{table*}
% \clearpage
\clearpage
\clearpage
\balance

\end{document}